\documentclass[10pt,journal]{IEEEtran}
\usepackage{amsmath,amssymb,amsfonts}
\usepackage{graphicx}
\usepackage{multirow}
\usepackage{cuted}
\usepackage{CJK}
\usepackage{indentfirst}
\usepackage{xcolor}
\usepackage{hyperref}
\usepackage{epstopdf}
\usepackage{bm}
\usepackage{cases}
\usepackage{subfigure}
\usepackage{caption}
\usepackage{bookmark}
\usepackage{makecell}

\usepackage{algorithmic}
\usepackage{algorithm}
\usepackage{array}
\usepackage{textcomp}
\usepackage{stfloats}
\usepackage{url}
\usepackage{verbatim}
\usepackage{cite}

\hypersetup{
colorlinks=true,
linkcolor=black,
citecolor=black
}

\begin{document}
\title{Joint Location and Velocity Estimation and Fundamental CRLB Analysis for Cell-Free MIMO-ISAC}
\author{Guoqing Xia, Pei Xiao, \textit{Senior Member, IEEE}, Qu Luo, \textit{Member, IEEE}, Bing Ji, \textit{Senior Member, IEEE}, Yue Zhang, \textit{Senior Member, IEEE}, and Huiyu Zhou.
\thanks{This work was supported in part by the U.K. Engineering and Physical Sciences Research Council under Grant EP/X013162/1. (Corresponding author:
Huiyu Zhou and Qu Luo.)
}

\thanks{Guoqing Xia and Bing Ji are with the School of Engineering, University of Leicester, LE1 7RH
Leicester, UK (e-mail: gx21@leicester.ac.uk; bing.ji@leicester.ac.uk).}
\thanks{Pei Xiao is with 5GIC $\&$ 6GIC, Institute for Communication Systems (ICS) of University of Surrey,
Guildford, GU2 7XH, UK (e-mail: p.xiao@surrey.ac.uk).}
\thanks{Qu Luo is with Institute for Communication Systems (ICS) of University of Surrey,
Guildford, GU2 7XH, UK (e-mail: q.u.luo@surrey.ac.uk).}
\thanks{Yue Zhang is with the Institute for Communication Systems and Measurement of China, Chengdu 610095, China (e-mail: zhangyue@icsmcn.cn).}
\thanks{Huiyu Zhou is with the School of Computing and Mathematical Sciences, University of Leicester, LE1 7RH
Leicester, UK (e-mail: hz143@leicester.ac.uk).}
}
\maketitle
\begin{abstract}

This paper presents a fundamental performance analysis of joint location and velocity estimation in a cell-free (CF) MIMO integrated sensing and communication (ISAC) system.  
Unlike prior studies that primarily rely on continuous-time signal models, we consider a more practical and challenging scenario in the discrete-time digital domain. 
Specifically, we first formulate a logarithmic likelihood function (LLF) and corresponding maximum likelihood estimation (MLE)  for both single- and multiple-target sensing.
Building upon the proposed LLF framework, closed-form Cramér-Rao lower bounds (CRLBs) for  joint location and velocity estimation are derived 
under deterministic, unknown, and spatially varying radar cross-section (RCS) models.  These CRLBs can serve as a fundamental performance metric to guide CF MIMO-ISAC system design. 
To enhance tractability, we also develop a class of simplified closed-form CRLBs, referred to as approximate CRLBs, along with a rigorous analysis of the conditions under which they remain accurate. Furthermore, we investigate how the sampling rate, squared effective bandwidth, and time width influence CRLB performance. For multi-target scenarios,  the concepts of safety distance and safety velocity are introduced to characterize the conditions under which the CRLBs converge to their single-target counterparts. Extensive simulations using orthogonal frequency division multiplexing (OFDM) and orthogonal chirp division multiplexing (OCDM) validate the theoretical findings and provide practical insights for CF MIMO-ISAC system design.

\end{abstract}
\section{Introduction}
\IEEEPARstart{I}{ntegrated} sensing and communication (ISAC) has emerged as a transformative technology for sixth-generation (6G) systems, enabling unified solutions for applications such as autonomous vehicles, smart cities, and remote healthcare \cite{Pang2024,Nuria2024}. Unlike conventional systems with separate communication and sensing modules, ISAC harnesses shared spectral and hardware resources to execute both functions simultaneously, offering substantial improvements in spectral efficiency, energy savings, and hardware utilization \cite{Liu2022survey, Xiong2023}. The multiple-input, multiple-output (MIMO) technology facilitates this integration by providing spatial diversity and multiplexing gains for both communication and sensing \cite{Liu2022, Haimovich2008}.

A particularly promising architecture for ISAC is cell-free (CF) MIMO, where a large number of geographically distributed access points (APs) are connected to a central processing unit \cite{Ganesan2021,Guo2022,Ammar2022Cell_free}. This architecture offers flexible coverage and rich spatial diversity, making it naturally suited for distributed ISAC systems. Recent research has started exploring CF MIMO-empowered ISAC systems for their potential to unify communication and sensing in a scalable and robust manner \cite{Behdad2022,Behdad2024, Ahmed2019, Huang2023, Ahmed2024, Weihao2024}.

Practical ISAC systems support a range of sensing tasks such as target detection \cite{Tajer2010,Fang2023,An2023}, and parameter estimation of radar cross-section (RCS) \cite{Dwivedi2018,Qi2022}, location \cite{Dianat2013,Park2015,Chen2023}, velocity \cite{He2010Velocity,Wang2023}, and angle \cite{Boyer2011,Liao2018}. These capabilities are essential for enabling applications such as autonomous navigation and environmental monitoring. In this context, robust and application-relevant performance metrics are essential—not only for assessing system effectiveness but also for guiding resource allocation and managing trade-offs between sensing and communication \cite{Liu2022survey,Xiong2023}. While unified metrics such as mutual information \cite{Ouyang2023MI} and Kullback-Leibler divergence \cite{Mohammad2023} have gained traction, their adaptation to specific sensing tasks remains non-trivial and is often oversimplified.
Therefore, we aim to bridge the gap by deriving theoretical performance bounds based on the Cramér–Rao lower bound (CRLB) for CF MIMO-ISAC systems, under practical and previously unexplored conditions.

Notably, CF MIMO shares structural similarities with distributed multi-static (DMS) MIMO radar \cite{Behdad2022, Ahmed2019, Huang2023, Ahmed2024, Behdad2024}. To maintain consistency with established radar literature, we adopt the term “DMS MIMO” for our sensing analysis.

\subsection{Related Works}
Conventional phased-array radars form narrow high-gain beams, focusing energy in specific directions. In contrast, MIMO radar systems leverage waveform diversity by transmitting orthogonal waveforms from independent antennas \cite{Li2007}. DMS MIMO radar extends this further by employing widely separated transmitters and receivers, taking advantage of spatial diversity to mitigate the impact of RCS scintillation, thus providing improved robustness in multipath environments and enhanced detection and estimation accuracy \cite{Fishler2006, Haimovich2008}.

Prior work on MIMO radar has extensively studied various sensing tasks, including target detection \cite{Tajer2010,Fang2023,An2023}, parameter estimation \cite{Dianat2013,Park2015,Chen2023,He2010Velocity,Wang2023}, waveform design \cite{Wang2014TGRS,Aldayel2016,Xue2024}, and theoretical performance bounds \cite{He2010, Chuanming2010, he2012noncoherent, He2016, Godrich2010, Ai2015, Godrich2012, Godrich2011}. Among these, CRLB-based analysis has emerged as a rigorous and widely adopted approach for quantifying estimation performance.
Based on RCS characteristics, two CRLB modeling approaches have been reported: one assuming a random RCS and the other assuming a deterministic RCS.

Under the random RCS-based assumption, RCSs are modeled as random variables, and their likelihood functions are derived by averaging over their joint distribution \cite{Chuanming2010, He2010, he2012noncoherent, He2016}. This approach allows for non-coherent processing and facilitates performance analysis when prior information is limited. Notably, \cite{He2010} derives CRLBs for joint location and velocity estimation and analyzes antenna deployment impacts. \cite{Chuanming2010} further extends this to multi-target scenarios, while \cite{He2016} introduces generalized and mismatched CRLBs to handle signal non-orthogonality and spatially correlated noise.

Under the deterministic RCS-based assumption, RCSs are considered to be deterministic, unknown and spatially diverse, requiring estimation alongside other parameters. This assumption is especially relevant in scenarios with slowly varying environments or no prior RCS information \cite{Godrich2010, Godrich2011, Godrich2012, Ai2015}. Under this model, CRLBs for location estimation have been derived for both coherent and non-coherent systems, with performance scaling with carrier frequency and effective bandwidth \cite{Godrich2010}. \cite{Ai2015} extends the results to multi-target scenarios, revealing the asymptotic convergence of individual CRLBs to their single-target counterparts. These insights have informed CRLB-based sensor selection and power allocation strategies \cite{Godrich2011, Godrich2012}, and have recently been adapted to CF MIMO-ISAC systems for location-based optimization \cite{Ahmed2019, Huang2023, Ahmed2024}.
It should be noted that existing analyses predominantly adopt continuous-time signal models and focus on single-parameter estimation.

\subsection{Motivations and Contributions}
While existing studies provide valuable theoretical foundations, several critical gaps remain. First, joint estimation of location and velocity under the deterministic RCS assumption in DMS MIMO radar and CF MIMO-ISAC systems remains largely unexplored. 
Second, despite the significance of discrete-time signal processing in practical radar systems, its impact on CRLB analyses, particularly the influence of sampling rate, has been inadequately addressed in the literature. Finally, most resource allocation strategies in DMS radar sensing (e.g., \cite{Godrich2011,feng2016fast}) or CF MIMO-ISAC systems (e.g., \cite{Behdad2022, Ahmed2019, Huang2023, Ahmed2024, Behdad2024}) focus on a single sensing objective, such as localization, due to lack of a joint multi-parameter sensing metric with simplified nonlinearity and nonconvexity.
 
This paper addresses these gaps by presenting a fundamental performance analysis of joint location and velocity estimation in CF MIMO-ISAC systems, grounded in a discrete-signal model under a deterministic RCS assumption. Our contributions are summarized as follows:

\begin{enumerate}
  \item[1)] We first derive the likelihood functions for the target parameters based on discrete received signals 
  in both single- and multiple-target sensing scenarios, emphasizing the critical role of the sampling rate. Building on this foundation, we derive, for the first time, the CRLBs for joint location and velocity estimation under the deterministic RCS assumption.
  \item[2)] To enhance the tractability of CRLB computation, we propose novel closed-form approximate CRLB expressions. We rigorously establish sufficient conditions to ensure their accuracy, offering both theoretical validation and practical relevance. These approximations enable efficient sensing performance evaluation, serving as useful metrics for cell-free MIMO-ISAC system design and resource allocation. Furthermore, we provide a comprehensive analysis of the trade-offs and behavior of the proposed (approximate) CRLBs, examining their sensitivity to signal-to-clutter-plus-noise ratio (SCNR), waveform parameters, sampling rates, and system configurations.
  \item[3)] We extend the CRLB framework to multi-target scenarios and analytically examine the coupling effects among targets. By establishing asymptotic relationships between single- and multiple-target CRLBs, we introduce the notions of \emph{safety distance} and \emph{safety velocity}, which define the spatial and velocity separations necessary for independent estimation performance among multiple targets.
  \item[4)] We validate the theoretical analyses through comprehensive simulations using widely adopted ISAC waveforms, i.e., orthogonal frequency division multiplexing (OFDM) and orthogonal chirp division multiplexing (OCDM). The proposed CRLBs are shown to align closely with the MSE performance of maximum likelihood estimators (MLE), indicating the accuracy of the derived (approximate) CRLBs.  
\end{enumerate}

The remainder of this paper is organized as follows.
Section \ref{Sec:syst model} presents the received signal model. Section \ref{sec:MLE} presents the discrete signal-based likelihood functions and discusses the MLE method. Section \ref{sec:CRLBs} provides a detailed derivation and analysis of the CRLB. Section \ref{sec:simulations} demonstrates the simulation results to validate the theoretical findings. Section \ref{sec:Conclusions} summarizes the conclusions of this work\footnote{To ensure that the hyperlinks between the main text and the appendices function correctly, we have also uploaded the full manuscript to arXiv, which is accessible via the following link: https://doi.org/10.48550/arXiv.2503.06766.}.

\emph{Notation}: $\mathbb{R}$ and $\mathbb{C}$ denote the field of real and complex numbers, respectively. Scalars are denoted by lower-case letters, vectors and matrices, respectively, by lower- and upper-case boldface letters. $\mathfrak{R}$ and $\mathfrak{T}$ are used to take the real part and the imaginary part, respectively. The conjugate, transpose, and conjugate transpose are denoted by $(\cdot)^*$, $(\cdot)^{\rm T}$, and $(\cdot)^{\rm H}$, respectively. $\mathbb{E}\{\cdot\}$ and $|\cdot|$ denote the mathematical expectation and element-wise modulus, respectively. ${\rm diag}\{\cdot\}$ generates a diagonal matrix by utilizing the input entries as the diagonal elements or diagonal blocks in the given order. $\|\cdot\|_2$ denotes the $l_2$ norm of a matrix. The expression $a\ll b$ means that $a$ is far less than $b$. ${\bm A}\succ   {\bm B}$ (${\bm A}\succeq   {\bm B}$) indicates that ${\bm A}-{\bm B}$ is a positive (semi-) definite matrix.  

\section{Signal Model}\label{Sec:syst model}
\begin{figure}
   \centerline
   {\includegraphics[width=0.23\textwidth]{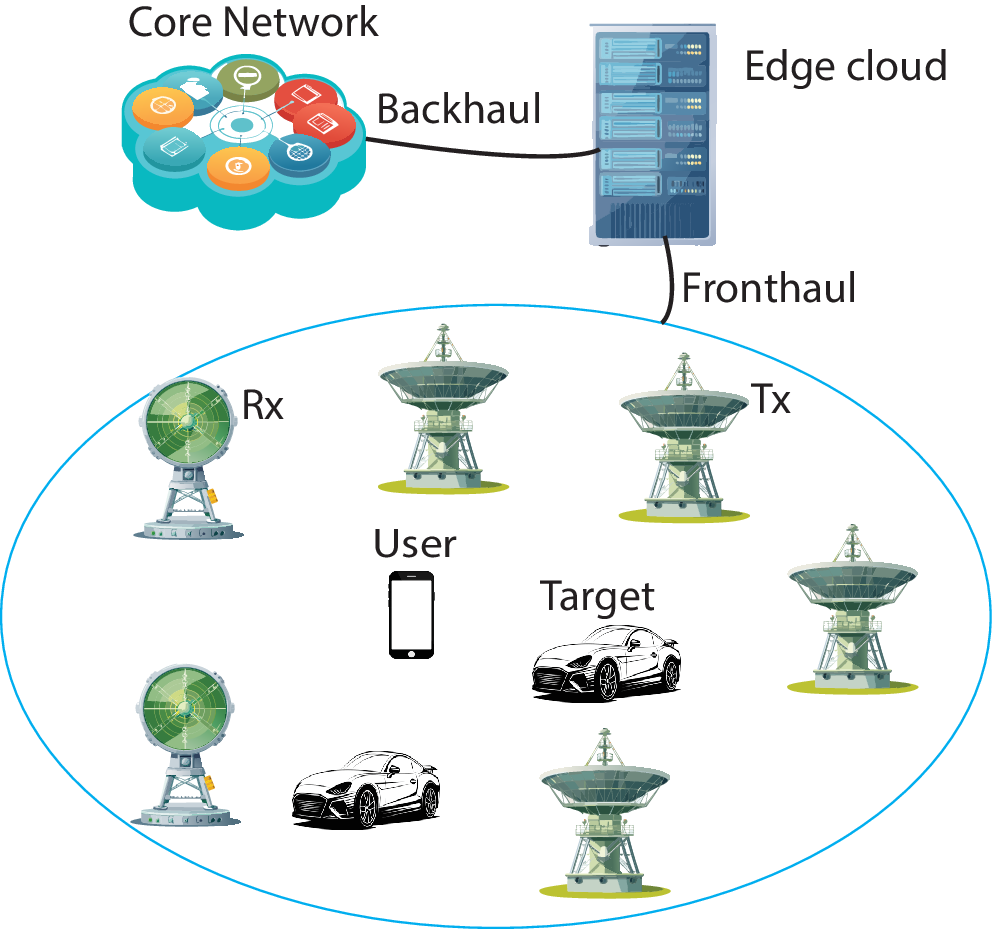}}
   \caption{CF MIMO system for ISAC.}\label{fig:MIMO_syst}
   \vspace{-0.3cm}
\end{figure}

As shown  in Fig. \ref{fig:MIMO_syst}, we consider a CF MIMO-ISAC system with widely separated APs. Each AP serves as an ISAC transmitter or a sensing receiver, similar to the setup in \cite{Behdad2022,Behdad2024}.  Assume that there are $N$ single-antenna  transmitters and $K$ single-antenna receivers, with the positions of the transmitter $n$ and receiver $k$ denoted as ${\bm l}_n = [x_n, y_n]^{\rm T}$ and ${\bm l}_k = [x_k, y_k]^{\rm T}$, respectively. All transmitters and receivers are scheduled and managed by the edge cloud and the core network, e.g., the fifth generation (5G) communication networks. Denote the location and velocity vectors of the $q$th sensing target as ${\bm l}_q = [x_q, y_q]^{\rm T}$ and ${\bm v}_{q} = [v_{x,q}, v_{y,q}]^{\rm T}$, respectively.

The Doppler frequency introduced by the relative radial velocity of a target versus any transmitter $n$ and receiver $k$, is computed as
\begin{equation}
\small
   f_{n,k,q} =  \dfrac{1}{\lambda}{\bm v}_{q}^{\rm T}\left(\frac{{\bm l}_n-{\bm l}_q}{\|{\bm l}_n-{\bm l}_q\|_2}+\frac{{\bm l}_k-{\bm l}_q}{\|{\bm l}_k-{\bm l}_q\|_2}\right), \label{eq:Doppler}
\end{equation}
where $\lambda$ denotes the wavelength.
Denote $\tau_{n,k,q}$ as the propagation delay between the $n$th transmitter and the $k$th receiver, as reflected by the $q$th target. Then, we have
\begin{subequations}
   \begin{align}
   \small
      \tau_{n,k,q}&=\tau_{n,q}+\tau_{k,q},\label{eq:trans delay}\\
      \tau_{n,q}&=\dfrac{\|{\bm l}_n-{\bm l}_q\|_2}{c},\label{eq:tx delay}\\
      \tau_{k,q}&=\dfrac{\|{\bm l}_k-{\bm l}_q\|_2}{c},\label{eq:rx delay}
   \end{align}
\end{subequations}   
where $\tau_{n,q}$ and $\tau_{q,k}$ represent the propagation delays from the $n$th transmitter to the $q$th target, and from the target to the $k$th receiver, respectively, and $c$ denotes the speed of light. For the single-target scenario, the subscript $q$ can be omitted.

A waveform $s_n(t)$,  with unit energy $\varepsilon_{s}=\int |{s}_{n}(t)|^2\,dt=1$,  is considered for both sensing and communication for the $n$th transmitter.  The baseband transmitted signal at transmitter $n$ is denoted as 
\begin{equation}
   {\bm x}_{n}(t) = \sqrt{P\rho_{n}}{b}_{n}\varsigma s_n(t), \label{eq:transmit sig sensing}
\end{equation}
where $P$ denotes the total available energy, $\rho_{n}$ is the energy allocation (EA) factor with $\sum_{n=1}^N\rho_n=1$, ${b}_{n}$ is the normalized transmit beamforming weight with $|{b}_{n}|^2=1$, and $\varsigma$ is the complex data symbol with $\mathbb{E} \{|\varsigma|^2\}=1$. Denote
${\bm \rho}=[\rho_1, \rho_2, \cdots, \rho_N]^{\rm T}$ by the vector of the EA coefficient of the $N$ transmitters. 

The received signal of the communication user is given by,
\begin{align}
   {r}_{u}(t) &= \sqrt{P}\sum_{n=1}^Ng_{n}{b}_{n}\sqrt{\rho_{n}}\varsigma s_n(t)+{z}(t),\label{eq:received sig commun}  
\end{align}
where $g_{n}$ is the channel fading between the transmitter $n$ and the user, and the interference-plus-noise ${z}(t)$ is assumed to be the additive circular Gaussian white with zero mean and variance $\sigma_{z}^2$. The diversity gain can be achieved by using the same $\varsigma$ on all transmitters. When mutually orthogonal waveforms $s_n(t)$ are employed, different symbols $\varsigma_n$ can be transmitted to realize multiplexing gain by allocating higher EA coefficients $\rho_n$ to the weaker channels characterized by $|g_{n}b_{n}|$.

For target-reflected reception, we denote the attenuation coefficient that combines both channel fading and radar RCS as $\alpha_{n,k,q}$. For simplicity, hereafter, we refer to the term ``RCS'' as encompassing both channel fading and radar RCS. Then, we define the RCS vector as ${\bm \alpha}_{n,k}\in \mathcal{C}^{Q\times 1}$. The Swerling-I sensing model is considered, i.e. the RCSs are assumed to be deterministic but unknown during the observation time \cite{van2001detection,Godrich2010,Godrich2011,Godrich2012,feng2016fast}.
The lowpass equivalent of the reflected signal from the $n$th transmitter by $Q$ targets received at receiver $k$ is given by\footnote{The transmitted signals over different transmitters are orthogonal and maintain approximate orthogonality over different delays and Doppler frequencies such that the signals contributed from different transmitters can be separated at each receiver \cite{Fishler2006, Haimovich2008,He2010,Chuanming2010,he2012noncoherent}.\\}
\begin{equation}
\begin{aligned}
   {r}_{n,k}(t) = \ &  {\bm \alpha}_{n,k}^{\rm T}{\bm y}_{n,k}(t)+z_{n,k}(t),\label{eq:received reflected sig}  
\end{aligned}
\end{equation}
where $z_{n,k}(t)$ denotes the clutter-plus-noise and ${\bm y}_{n,k}(t)$ is the loss-free and noise-free received signal vector with its $q$ element being,
\begin{align}
{y}_{n,k,q}(t)=\sqrt{P\rho_{n}}b_n\varsigma s_{n}(t-\tau_{n,k,q})e^{j2\pi f_{n,k,q}t}.
\end{align}
Note that clutters can initially be mitigated by Doppler processing or space-time adaptive processing \cite{Melvin2013ModernRadar} and whitened through a whitening filter \cite{He2010}. It is reasonable to assume the clutter-plus-noise to be additive white Gaussian noise with zero mean and variance $\sigma_{z}^2$ \cite{Haimovich2008,He2010Homogeneous,Wang2011Nonhomogeneous,He2010,Chuanming2010,he2012noncoherent}. Define the SCNR as $\delta=\frac{P}{T_{\rm eff}\sigma_{z}^2}$ with $T_{\rm eff}$ being the effective pulse time width.

\section{ Maximum Likelihood Estimation Based on Discrete Received Signal}\label{sec:MLE}
In this section, we first derive the discrete signal-based logarithmic likelihood function (LLF).
Building upon the derived LLF, we introduce the MLE methods for joint location and velocity estimation in both single- and multiple-target scenarios. Finally, the significance of the sampling rate is analyzed through MLE estimation. 

To begin with, define the unknown parameter of the $q$th target in multiple-target sensing scenario as
\begin{align}
   {\bm \alpha}_q \triangleq\ & [\alpha_{1,1,q},\alpha_{1,2,q},\cdots,\alpha_{N,K,q}]= {\bm \alpha}_{{\rm Re},q}+j{\bm \alpha}_{{\rm Im},q}, \\
   {\bm \theta}_q\triangleq \ &[{\bm l}_q^{\rm T}, {\bm v}_q^{\rm T}, {\bm \alpha}^{\rm T}_{{\rm Re},q},{\bm \alpha}^{\rm T}_{{\rm Im},q}]^{\rm T},\\
  {\bm \phi}_q\triangleq \ &[\tau_{1,1,q},\tau_{1,2,q},\cdots,\tau_{N,K,q},f_{1,1,q},\cdots,f_{N,K,q},\notag\\
  &{\bm \alpha}^{\rm T}_{{\rm Re},q},{\bm \alpha}^{\rm T}_{{\rm Im},q}]^{\rm T},
\end{align}
where $f_{n,k,q}$ and $\tau_{n,k,q}$ are given in \eqref{eq:Doppler} and \eqref{eq:trans delay}, respectively. Then, the parameter vectors to be estimated for multiple targets are denoted by
${\bm \theta}_{\rm MT}=[{\bm \theta}_1^{\rm T},{\bm \theta}_2^{\rm T},\cdots,{\bm \theta}_Q^{\rm T}]^{\rm T}$ and 
${\bm \phi}_{\rm MT}=[{\bm \phi}_1^{\rm T},{\bm \phi}_2^{\rm T},\cdots,{\bm \phi}_Q^{\rm T}]^{\rm T}$, respectively.
Similarly, define the unknown parameter vector for single-target sensing as
\begin{align}
   {\bm \alpha}&\triangleq [\alpha_{1,1},\alpha_{1,2},\cdots,\alpha_{N,K}] = {\bm \alpha}_{\rm Re}+j{\bm \alpha}_{\rm Im}, \\
   {\bm \theta}&\triangleq [{\bm l}^{\rm T}, {\bm v}^{\rm T}, {\bm \alpha}^{\rm T}_{\rm Re},{\bm \alpha}^{\rm T}_{\rm Im}]^{\rm T}\in\mathbb{R} ^{(4+2NK)\times 1},\\
  {\bm \phi}&\triangleq [\tau_{1,1},\tau_{1,2},\cdots,\tau_{N,K},f_{1,1},\cdots,f_{N,K},{\bm \alpha}^{\rm T}_{\rm Re},{\bm \alpha}^{\rm T}_{\rm Im}]^{\rm T}.
\end{align}
Note that the subscripts related to target indexes are omitted hereafter for the single-target sensing scenario. 
\subsection{discrete signal-based LLF}\label{sec:discrete LLF}
For practical implementation, we now provide LLFs based on discrete signals for single- and multiple-target sensing. A narrowband sensing waveform is considered to allow multiple scattered signals from an extended target to be treated as a whole, with the same time delay \cite{Godrich2010}.

The discrete received signal from the $n$th transmitter at the $k$th receiver is expressed as  
\begin{equation}
\begin{aligned}
            r_{n,k}(i)& =   r_{n,k}(t) \vert _{t = it_s},\\
            &= {\bm \alpha}_{n,k}^{\rm T}{\bm y}_{n,k}(i)+z_{n,k}(it_s), \label{eq:simplified reflected sig nk}
\end{aligned}
\end{equation}
where $r_{n,k}(t)$ is given in \eqref{eq:received reflected sig}, $t_{\rm s}=1/f_s$ is the sampling interval with $f_s$ denoting the sampling rate, and ${\bm y}_{n,k}(i)$ is the discrete loss-free and noise-free received signal vector with its $q$th element being ${y}_{n,k,q}(i)=y_{n,k,q}(t) \vert _{t = it_s}$.

With the number of samples $S=f_sT_{\rm eff}$ at a given sampling rate $f_s$, we define the discrete received signal vector as 
\begin{align}
   {\bm r}=[{\bm r}_{1,1}^{\rm T}, {\bm r}_{1,2}^{\rm T},\cdots,{\bm r}_{1,K}^{\rm T},{\bm r}_{2,1}^{\rm T},\cdots,{\bm r}_{N,K}^{\rm T}]^{\rm T}
\end{align} 
with ${\bm r}_{n,k}=[r_{n,k}(1),r_{n,k}(2),\cdots, r_{n,k}(S)]$.
Then, the LLF for multiple-target sensing can be given by 
\begin{equation}
\small
\begin{aligned}
   \log p({\bm r}|{\bm \phi}_{\rm MT})=\ &\sum_{n=1}^N\sum_{k=1}^K\log \prod_{i=1}^{S}p(r_{n,k}(i)|{\bm \phi}),\\
   =\ &\sum_{n=1}^N\sum_{k=1}^K\log e^{-\frac{1}{\sigma_{z}^2}\sum_{i=1}^{S}|{r}_{n,k}(i)-{\bm \alpha}_{n,k}^{\rm T}{\bm y}_{n,k}(i)|^2}\\
   =\ & \frac{2}{\sigma_{z}^2}\sum_{n=1}^N\sum_{k=1}^K\mathfrak{R}\left\{ {\bm \alpha}_{n,k}^{\rm H}\sum_{i=1}^{S}{r}_{n,k}(i){\bm y}_{n,k}^*(i)\right\} \\
   -&\frac{1}{\sigma_{z}^2}\sum_{n=1}^N\sum_{k=1}^K{\bm \alpha}_{n,k}^{\rm T}\sum_{i=1}^{S}{\bm y}_{n,k}(i){\bm y}_{n,k}^{\rm H}(i){\bm \alpha}_{n,k}^{*}.
\end{aligned}\label{eq:LF multiple}
\end{equation}

For single-target sensing, the discrete received signal can be expressed as
\begin{equation}
   \small
   r_{n,k}(i) = \alpha_{n,k} \underbrace{\sqrt{P\rho_{n}}b_n\varsigma  s_{n}(it_{\rm s}-\tau_{n,k})e^{j2\pi f_{n,k}it_{\rm s}}}_{y_{n,k}(i)}+z_{n,k}(it_{\rm s}), 
\end{equation}
Similar to \eqref{eq:LF multiple}, the single-target LLF $\log p({\bm r}|{\bm \phi})$ is expressed by
\begin{equation}
   \small
\begin{aligned}
   \log p({\bm r}|{\bm \phi})=\ &\frac{2}{\sigma_{z}^2}\sum_{n=1}^N\sum_{k=1}^K\mathfrak{R}\left\{ {\alpha}_{n,k}\sum_{i=1}^{S}{r}_{n,k}^{*}(i){y}_{n,k}(i)\right\} \\ 
  - &  \frac{1}{\sigma_{z}^2}\sum_{n=1}^N\sum_{k=1}^KP\rho_{n}|{\alpha}_{n,k}\varsigma |^2\sum_{i=1}^{S}|s_{n}(it_{\rm s}  \! - \! \tau_{n,k})|^2.
\end{aligned}\label{eq:LF}
\end{equation}
It can be further approximated according to the integral definition as
\begin{equation}
   \small
   \begin{aligned}
   \log p({\bm r}|{\bm \phi})&\approx \frac{2}{\sigma_{z}^2}\sum_{n=1}^N\sum_{k=1}^Kf_s\mathfrak{R}\left\{ {\alpha}_{n,k}\int{r}_{n,k}^{*}(t){y}_{n,k}(t)\, dt\right\}\\ 
   &-\frac{1}{\sigma_{z}^2}\sum_{n=1}^N\sum_{k=1}^Kf_sP\rho_{n}|{\alpha}_{n,k}\varsigma |^2.
\end{aligned}\label{eq:LLF}
\end{equation}





\subsection{Single-target MLE}\label{sec:ST MLE}
We now introduce the single-target MLE. First rewrite the discrete signal-based LLF $\log p({\bm r}|{\bm \theta})$ in \eqref{eq:LLF} as 
\begin{equation}
\small
  \begin{aligned}
   \log p({\bm r}|{\bm \theta})&\approx \frac{2}{\sigma_{z}^2}\sum_{n=1}^N\sum_{k=1}^K\mathfrak{R}\left\{ {\alpha}_{n,k}^{*}\sum_{i=1}^S{r}_{n,k}(i)\tilde{y}_{n,k}^*(i)\right\}\\ 
   &-\frac{1}{\sigma_{z}^2}\sum_{n=1}^N\sum_{k=1}^Kf_sP\rho_{n}|{\alpha}_{n,k}\varsigma |^2, \label{eq:LLF1}
\end{aligned}  
\end{equation}
with $\tilde{y}_{n,k}(i)=\sqrt{P\rho_{n}}b_ns_{n}(it_{\rm s}-\tilde\tau_{n,k})e^{j2\pi \tilde{f}_{n,k}it_{\rm s}}$ denoting the matching waveform onto ${y}_{n,k}(i)$ generated by using the given location (delay) and velocity (Doppler frequency) parameters.  

A necessary condition for MLE is given by $\dfrac{\partial{\log p({\bm r}|{\bm \theta})}}{\partial{{\bm \theta}^*}}={\bm 0}$ \cite{John2006Chapter4,Lehmann2006,Godrich2010}. By solving $\dfrac{\partial{\log p({\bm r}|{\bm \theta})}}{\partial{{\bm \alpha}^*}}={\bm 0}$, we have the MLE solution of ${\alpha}_{n,k}$, 
$\hat{\alpha}_{n,k}=\frac{\sum_{i=1}^S{r}_{n,k}(i)\tilde{y}_{n,k}^*(i)}{2P\rho_{n}f_s|\varsigma |^2}$. Replacing ${\alpha}_{n,k}$ in \eqref{eq:LLF1} with $\hat{\alpha}_{n,k}$ and assuming the same EA  for each $n$ (i.e., $\rho_n=1/N$), we can obtain the MLE function regarding both the location and velocity parameters as
\begin{equation}
   \small
   \log p({\bm r}|\{{\bm l}, {\bm v}\})\approx \sum\nolimits_{n=1}^N\sum\nolimits_{k=1}^K\bigg|\sum\nolimits_{i=1}^S{r}_{n,k}(i)\tilde{y}_{n,k}^*(i)\bigg|^2, \label{eq:modf LLF}
\end{equation}
where insignificant constants have been omitted for simplicity. 
By maximizing \eqref{eq:modf LLF}, one can perform a four-dimensional grid search to jointly estimate the location and velocity \cite{Godrich2010,He2010}. 

Furthermore, the LLF in \eqref{eq:modf LLF} can be further written by using \eqref{eq:simplified reflected sig nk} as
\begin{equation}
\small
   \log p({\bm r}|\{{\bm l}, {\bm v}\})\approx \sum_{n=1}^N\sum_{k=1}^K|{\alpha}_{n,k}|^2\bigg|\sum_{i=1}^S{y}_{n,k}(i)\tilde{y}_{n,k}^*(i)\bigg|^2, \label{eq:AF discrete}
\end{equation}
where we have used $\sum_{i=1}^Sz_{n,k}(i)\tilde{y}_{n,k}^*(i)\approx 0$ in \eqref{eq:AF discrete} for white Gaussian clutter-plus-noise $z_{n,k}(i)$ with assumed sufficient number of samples $S$. {\it Equation \eqref{eq:AF discrete} defines the discrete ambiguity function (AF), indicating the dependence of MLE performance on the inherent AF when $S$ is large enough.} 

\subsection{Multiple-target MLE} \label{sec:Multi-tar MLE}
We now study the MLE method for sensing multiple targets. 
First simplify the multiple-target LLF from \eqref{eq:LF multiple} for a large sampling rate, i.e.,
\begin{equation}
   \small
  \begin{aligned}
   \log p({\bm r}|{\bm \theta}_{\rm MT})\approx\ &\frac{2}{\sigma_{z}^2}\sum\nolimits_{n=1}^N\sum\nolimits_{k=1}^K\mathfrak{R}\left\{ {\bm \alpha}_{n,k}^{\rm H}{\bm d}_{n,k}\right\}\\ 
   &-\frac{1}{\sigma_{z}^2}\sum\nolimits_{n=1}^N\sum\nolimits_{k=1}^Kf_s{\bm \alpha}_{n,k}^{\rm T}\tilde{\bm Y}_{n,k}{\bm \alpha}_{n,k}^{*},\label{eq:LLF multiple discrete}
\end{aligned}  
\end{equation}
where ${\bm d}_{n,k}\triangleq\sum_{i=1}^S{r}_{n,k}(i)\tilde{\bm y}_{n,k}^*(i)$ is the discrete cross-correlation vector, and $\tilde{\bm Y}_{n,k} \triangleq  \int \tilde{\bm y}_{n,k}(t)\tilde{\bm y}_{n,k}^{\rm H}(t)\, dt$ denotes the cross-correlation matrix over $Q$ targets. Note that $\tilde{\bm y}_{n,k}(i)$ is the waveform that matches ${\bm y}_{n,k}(i)$ using the given locations and velocities of multiple targets.
By solving $\dfrac{\log p({\bm r}|{\bm \theta})}{\partial{{\bm \alpha}_{n,k}^*}}={\bm 0}$, we have $\hat{\bm \alpha}_{n,k}=\frac{1}{2f_s}(\tilde{\bm Y}_{n,k}^{*})^{-1}{\bm d}_{n,k}$. By replacing ${\bm \alpha}_{n,k}$ in \eqref{eq:LLF multiple discrete} with $\hat{\bm \alpha}_{n,k}$, we have the simplified MLE function as follows: 
\begin{equation}
   \small
   \log p({\bm r}|\{{\bm l}_q, {\bm v}_q\}_{q=1,\cdots,Q})\approx \sum\nolimits_{n=1}^N\sum\nolimits_{k=1}^K {\bm d}_{n,k}^{\rm H}\tilde{\bm Y}_{n,k}^{-1}{\bm d}_{n,k}.\label{eq:LLF multiple discrete simplified}
\end{equation}

It should be noted that the proposed simplified MLE functions, i.e., \eqref{eq:LLF multiple discrete} and \eqref{eq:LLF multiple discrete simplified}, make the MLE calculation feasible. 
With the proposed discrete signal-based MLE problem, existing estimation schemes, such as compressed sensing and relaxation-based approaches can be leveraged for low-complexity and accuracy estimations \cite{Zhang2022Survey}. A detailed discussion of this aspect is beyond the scope of this paper. 

We now investigate the evolution trend of multiple-target MLE towards corresponding single-target ones regarding the relative delay and relative Doppler frequency between one another. First, define the relative time delay of target $q$ and $l$ over transmitter $n$ and receiver $k$ as
$\tau_{n,k,q,l}\triangleq\tau_{n,k,q}-\tau_{n,k,l}$,
and the relative Doppler frequency as $f_{n,k,q,l}\triangleq f_{n,k,q}-f_{n,k,l}$.
The time delay resolution, $\tau_{\rm r}$, and the Doppler frequency resolution, $f_{\rm r}$, represent the minimum time delay and Doppler frequency for which the correlation between the transmit signal $s_n(t)$ and its corresponding time-frequency delayed counterpart $s_n(t-\tau_{\rm r})e^{j2\pi f_{\rm r}t}$ falls below a threshold\footnote{We here only present a qualitative definition, and in this paper the selection of the threshold needs to enable the decoupling of two targets with given relative delay or Doppler frequency \cite{Ai2015}.}. We readily have $T_{\rm eff}\geq \tau_{\rm r}$ and $B_{\rm eff}\geq f_{\rm r}$ where $B_{\rm eff}$ is the effective signal bandwidth. When targets are widely separated in delay-Doppler domain such that $|\tau_{n,k,q,l}|\geq \tau_{\rm r}$ or $|f_{n,k,q,l}|\geq f_{\rm r}$, we will have $|\tilde{Y}_{n,k,q,l}|/\tilde{Y}_{n,k,q,q}\rightarrow 0$. More details can be found in Appendix \ref{sec:CRLB trend}. In this case, the cross-correlation matrix can be approximated as a diagonal matrix, i.e., $\tilde{\bm Y}_{n,k}\approx P\rho_n{\bm I}_Q$, as $\tilde{Y}_{n,k,q,q}= P\rho_n$, $\forall q$. With a uniform EA, i.e, $\rho_n=1/N$, $\forall n$, \eqref{eq:LLF multiple discrete simplified} can be simplified as 
\begin{equation}
\small
   \log p({\bm r}|\{{\bm l}_q, {\bm v}_q\}_{q=1,\cdots,Q})\approx \sum_{q=1}^Q\sum_{n=1}^N\sum_{k=1}^K \bigg|\sum_{i=1}^S{r}_{n,k}(i)\tilde{y}_{n,k,q}^*(i)\bigg|^2.\label{eq:MLE function}
\end{equation}

The maximization of \eqref{eq:MLE function} can be achieved by maximizing the following $Q$ sub-functions, given by for each $q$
\begin{equation}
\small
   \log p({\bm r}|\{{\bm l}_q, {\bm v}_q\})\approx \sum_{n=1}^N\sum_{k=1}^K \bigg|\sum_{i=1}^S{r}_{n,k}(i)\tilde{y}_{n,k,q}^*(i)\bigg|^2,\label{eq:MLE sub function}
\end{equation}
which has the similar form with that of \eqref{eq:modf LLF}. When $|\tau_{n,k,q,l}|\geq \tau_{\rm r}$, we also have ${y}_{n,k,l}(t)\tilde{y}_{n,k,q}^*(t)\approx 0$. Therefore, \eqref{eq:MLE sub function} can be eventually simplified as
\begin{equation}
\small
   \log p({\bm r}|\{{\bm l}_q, {\bm v}_q\})\approx \sum_{n=1}^N\sum_{k=1}^K \bigg|\alpha_{n,k,q}\sum_{i=1}^S{y}_{n,k,q}(i)\tilde{y}_{n,k,q}^*(i)\bigg|^2.\label{eq:MLE sub function q}
\end{equation}

{\it Therefore, the sensing performance of one target can remain unaffected by other targets and the target-wise MLE can be solved using \eqref{eq:MLE sub function}, provided that the targets are sufficiently separated in both the delay and Doppler domains.} This conclusion will be explored in greater detail in terms of the theoretical performance analysis presented in Appendix \ref{sec:CRLB trend}.

\section{Fundamental Analysis of  Cramér-Rao Lower Bounds}\label{sec:CRLBs}
In this section, we first derive the CRLBs for joint location and velocity estimation in both single-target and multiple-target scenarios. Then, we provide an in-depth analysis of the CRLB in the DMS MIMO sensing systems\footnote{Similar to \cite{Sakhnini2022}, this paper primarily concentrates on the sensing performance analysis within the CF MIMO-ISAC framework. A comprehensive investigation of the entire ISAC system, including joint communication-sensing optimization, is left for future work due to space limitations.}.  

\subsection{The derived CRLB}
We now demonstrate the CRLBs for joint location and velocity estimation
in both single- and multiple-target sensing scenarios.
Before delving into the details, define the squared effective bandwidth (SEBW) and squared effective pulse time width (SETW) as
\begin{equation}
   \bar{f_n^2}=\frac{1}{4\pi^2}\int \bigg|\frac{d{s_{n}(t-\tau_{n,k})}}{d{\tau_{n,k}}}\bigg|^2\, dt=\int f^2|S_n(f)|^2\, df,\label{eq:square bandwidth}
\end{equation}
and
\begin{align}
   \bar{t_n^2}=\int t^2|s_n(t)|^2\, dt, \label{eq:square time}
\end{align}
respectively. The cross-term between time and frequency is further defined as 
\begin{align}
\sigma_{tf}=\int{ts_{n}^*(t-\tau_{n,k})}\frac{d{s_{n}(t-\tau_{n,k})}}{d{\tau_{n,k}}}\, dt.  \label{eq:cross term}
\end{align} 
We further define the average frequency as 
\begin{align}
\bar{f}_n=\int f|S_n(f)|^2\ df, \label{eq:average frequency}
\end{align}
and average time as
\begin{align}
    \bar{t}_n=\int t|s_n(t)|^2\ dt.  \label{eq:average time}
\end{align}
\subsubsection{Single-Target CRLB}\label{sec:ST CRLB}
The $(4+2NK)\times (4+2NK)$ single-target CRLB matrix is defined as the inversion of the FIM for ${\bm \theta}$,  i.e., 
\begin{equation}
   \small
{\bm C}\triangleq \left({\frac{\partial {\bm \phi}}{\partial{\bm \theta}}{\bm J}({\bm \phi})\left(\frac{\partial {\bm \phi}}{\partial{\bm \theta}}\right)^{\rm T}}\right)^{-1}.\label{eq:CRLB LV single}
\end{equation}
where ${\bm J}({\bm \phi})$ is the FIM for ${\bm \phi}$,  
\begin{equation}
   \small
{\bm J}({\bm \phi})\triangleq -\mathbb{E}_{{\bm r}}\left\{\frac{\partial^2 \log{p({{\bm r}}|{\bm \phi})}}{\partial{\bm \phi}{\partial{\bm \phi}^{\rm T}}}\right\},\label{eq:FIM_phi}
\end{equation}
with $p({\bm r}|{\bm \theta})$ and $p({\bm r}|{\bm \phi})$ denoting the likelihood functions of vector ${\bm r}$ conditioned on ${\bm \theta}$ and ${\bm \phi}$, respectively. 

To start with, the partial derivative $\frac{\partial {\bm \phi}}{\partial {\bm \theta}}$ is readily given by 
\begin{align}
   \frac{\partial {\bm \phi}}{\partial {\bm \theta}} = {\rm diag}\{{\bm \aleph},{\bm I}_{2NK}\},\label{eq:FO-phi-theta}
\end{align}
with ${\bm \aleph}$ characterizing the geometric spread of the sensing system with its expression given by
\begin{subequations}
    \setlength{\arraycolsep}{2pt}
\begin{align}
   {\bm \aleph} &= \begin{bmatrix}
      {\bm \aleph}_{11}&{\bm \aleph}_{12}\\{\bm O} &{\bm \aleph}_{22}
   \end{bmatrix}\label{eq:GeoSpread_mat}\\
   &=\begin{bmatrix}
\beta_{1,1}&\beta_{1,2}&\cdots&\beta_{N,K}&\eta_{1,1}&\eta_{1,2}&\cdots&\eta_{N,K}\\\zeta_{1,1}&\zeta_{1,2}&\cdots&\zeta_{N,K}&\kappa_{1,1}&\kappa_{1,2}&\cdots&\kappa_{N,K}\\0&0&\cdots&0&\xi_{1,1}&\xi_{1,2}&\cdots&\xi_{N,K}\\0&0&\cdots&0&\varrho_{1,1}&\varrho_{1,2}&\cdots&\varrho_{N,K}
   \end{bmatrix}, \label{eq:FOPD_mat}
\end{align}
\end{subequations}
where the element-wise definitions are provided by \eqref{eq:GeoSpread para} in Appendix \ref{sec: proof of theorem 5}. 

Although the CRLB has been extensively studied in existing works, deriving a closed-form expression, based on the proposed discrete signal-based LLFs and the deterministic RCS assumption, proves to be considerably challenging and far from straightforward. Through some mathematical manipulations, we derive the CRLB matrix in \eqref{eq:CRLB LV single} as follows
\begin{align}
   {\bm C} = \begin{bmatrix}
      {\bm \aleph}{\bm \varPhi} {\bm \aleph}^{\rm T}&{\bm \aleph}{\bm \varPsi}\\{\bm \varPsi}^{\rm T}{\bm \aleph}^{\rm T} &{\bm F}
      \end{bmatrix}^{-1},
\end{align}
where we define
\begin{align}
   {\bm \varPhi} \triangleq  \begin{bmatrix}{\bm A}&{\bm B}\\{\bm B} &{\bm D}\end{bmatrix},
\end{align}  
and 
\begin{align}
   {\bm \varPsi} \triangleq  \begin{bmatrix}{\bm G}\\{\bm E}\end{bmatrix}.
\end{align} 
According to the blockwise matrix inversion property in \cite{MatrixCookbook2012}, the joint location and velocity CRLB matrix, i.e., the upper $4\times 4$ block of ${\bm C}$ is given by  
\begin{align}
   {\bm C}_{\rm LV} = 
      ({\bm \aleph}{\bm \varPhi} {\bm \aleph}^{\rm T}-{\bm \aleph}{\bm \varPsi}{\bm F}^{-1}{\bm \varPsi}^{\rm T}{\bm \aleph}^{\rm T}
      )^{-1}. \label{eq:CRLB LV}
\end{align}

The computation of ${\bm C}_{\rm LV}$ in \eqref{eq:CRLB LV} is detailed as follows. 
Firstly, ${\bm A}$, ${\bm B}$ and ${\bm D}$ are diagonal matrices with their $((n-1)K+k)$th diagonal elements respectively given by
\begin{subequations}
   \begin{align}
   a_{n,k}&=\frac{8\pi^2|\alpha_{n,k}|^2f_sP\rho_{n}\bar{f_n^2}}{\sigma_{z}^2},\label{eq:simplified SO-PD delay}\\
   b_{n,k}&=\frac{4\pi |\alpha_{n,k}|^2f_sP\rho_{n}}{\sigma_{z}^2}\mathfrak{T}\{\sigma_{tf}\},\label{eq:simplified SO-PD Delay_Doppler}\\
   d_{n,k}&\approx\frac{8\pi^2|\alpha_{n,k}|^2f_sP\rho_{n}\bar{t_n^2}}{\sigma_{z}^2},\label{eq:simplified SO-PD Doppler}
   \end{align}
\end{subequations}
where the approximation  in \eqref{eq:simplified SO-PD Doppler}  holds when $\tau_{n,k}\ll  T_{\rm eff}$.

Additionally, ${\bm G}=[{\bm G}^{\rm Re}, {\bm G}^{\rm Im}]$ and ${\bm E}=[{\bm E}^{\rm Re},{\bm E}^{\rm Im}]$ contain the second-order derivatives regarding the delay (or Doppler frequency) and RCS, where ${\bm G}^{\rm Re}$, ${\bm G}^{\rm Im}$, ${\bm E}^{\rm Re}$ and ${\bm E}^{\rm Im}$ are diagonal matrices, with their $((n-1)K+k)$th diagonal elements respectively computed by 
\begin{subequations}
\begin{align}
g^{\rm Re}_{n,k}&=-\frac{2f_sP\rho_n}{\sigma_{z}^2}\mathfrak{R}\left\{\alpha^*_{n,k}\mu_{1,n,k}\right\},\label{eq:g re}\\
g^{\rm Im}_{n,k}&=\frac{2f_sP\rho_n}{\sigma_{z}^2}\mathfrak{T}\left\{\alpha^*_{n,k}\mu_{1,n,k}\right\},\\
e^{\rm Re}_{n,k}&=-\frac{2f_sP\rho_n}{\sigma_{z}^2}\mathfrak{R}\{\alpha^*_{n,k}\mu_{2,n,k}\},\\
e^{\rm Im}_{n,k}&=\frac{2f_sP\rho_n}{\sigma_{z}^2}\mathfrak{T} \left\{\alpha^*_{n,k}\mu_{2,n,k}\right\},\label{eq:g im}
\end{align}
\end{subequations}
where $\mu_{1,n,k}$ is the first moment of the energy spectrum
\begin{align}
\mu_{1,n,k}&\triangleq \int s_n^*(t-\tau_{n,k})\frac{\partial{s_n(t-\tau_{n,k})}}{\partial{\tau_{n,k}}}\, dt\notag\\
&=-j2\pi \bar{f}_n,
\end{align}
and $\mu_{2,n,k}$ is the first moment of the squared magnitude of the complex envelope
\begin{align}
   \mu_{2,n,k}&\triangleq j2\pi\int t|s_n(t-\tau_{n,k})|^2\, dt \notag\\
   &=j2\pi(\bar{t}_n+\tau_{n,k}).\label{eq:FO moment squared mag}
\end{align}

The final block matrix ${\bm F}$ is related to the RCS estimation, given by ${\bm F}={\rm diag}\{{\bm F}^{\rm Re},{\bm F}^{\rm Im}\}$
where the $((n-1)K+k)$th diagonal element of corresponding diagonal matrices ${\bm F}^{\rm Re}$ and ${\bm F}^{\rm Im}$ are respectively given by 
\begin{equation}
   \begin{cases}  
f^{\rm Re}_{n,k} =2f_sP\rho_n/\sigma_z^2,\\
f^{\rm Im}_{n,k} =2f_sP\rho_n/\sigma_z^2.
\end{cases}\label{eq:F elements}
\end{equation}
Note that ${\bm F}$ is a positive definite diagonal matrix. 

To make the CRLB calculation tractable and reduce computational complexity, an approximate—albeit looser—single-target CRLB is derived in \textbf{Theorem 1}, along with general sufficient and necessary conditions to ensure its tightness. Building on this, a waveform-dependent sufficient condition is further established in \textbf{Theorem 2}. Before presenting the detailed analysis, we first introduce a supporting lemma.

\textbf{Lemma 1}. \emph{For a symmetric matrix ${\bm \Xi}\succ 0$ and a symmetric matrix ${\bm \Theta}\succeq 0$ satisfying ${\bm \Xi}\succ {\bm \Theta}$, we always have ${\bm \Xi}^{-1}\preceq  ({\bm \Xi}-{\bm \Theta})^{-1}$ and $[{\bm \Xi}^{-1}]_{ii}\leq [({\bm \Xi}-{\bm \Theta})^{-1}]_{ii}$, where $[\cdot]_{ii}$ denotes the $i$th diagonal element of a matrix.}

\emph{Proof.} Firstly, symmetric matrix ${\bm \Theta}\succeq 0$ can be decomposed as ${\bm \Theta}={\bm U}{\bm U}^{\rm T}$ \cite{MatrixCookbook2012}. Define ${\bm M}=\begin{bmatrix}{\bm \Xi}&{\bm U}\\{\bm U}^{\rm T}&{\bm I}\end{bmatrix}$ with ${\bm I}$ being an identity matrix. According to the Schur complement conditions for positive definiteness \cite{Zhang2005Schur_complement}, we have,
\begin{align}
   &{\bm M}\succ 0\Leftrightarrow {\bm \Xi}\succ 0\ \&\ {\bm M}/{\bm I}={\bm \Xi}-{\bm \Theta}\succ 0\notag\\
   &\Leftrightarrow {\bm I}\succ 0\ \&\ {\bm M}/{\bm \Xi}={\bm I}-{\bm U}^{\rm T}{\bm \Xi}^{-1}{\bm U}\succ 0 ,
\end{align}
where ${\bm M}/{\bm I}$ and ${\bm M}/{\bm \Xi}$ are the Schur complement matrices of ${\bm M}$. Thus, ${\bm I}-{\bm U}^{\rm T}{\bm \Xi}^{-1}{\bm U}$ is symmetric and positive definite.
Then, according to the matrix inversion lemma \cite{MatrixCookbook2012}, $({\bm \Xi}-{\bm \Theta})^{-1}$ is simplified as  
\begin{equation}
   \small
   ({\bm \Xi}-{\bm \Theta})^{-1} = 
      {\bm \Xi}^{-1}+{\bm \Xi}^{-1}{\bm U}({\bm I}-{\bm U}^{\rm T}{\bm \Xi}^{-1}{\bm U})^{-1}{\bm U}^{\rm T}{\bm \Xi}^{-1}.  \label{eq:matrix inversion}
\end{equation} 
Since ${\bm \Xi}$ is symmetric and positive definite, its inverse ${\bm \Xi}^{-1}$ is likewise symmetric and positive definite. So do ${\bm I}-{\bm U}^{\rm T}{\bm \Xi}^{-1}{\bm U}$ and its inverse. Therefore, the second term on the right-hand side of \eqref{eq:matrix inversion} is a symmetric and positive semi-definite matrix such that ${\bm \Xi}^{-1}\preceq  ({\bm \Xi}-{\bm \Theta})^{-1}$. As a positive (semi-) definite symmetric matrix has positive (non-negative) diagonal elements, we further have $[{\bm \Xi}^{-1}]_{ii}\leq [({\bm \Xi}-{\bm \Theta})^{-1}]_{ii}$, $\forall{i}$. The equality holds in both ${\bm \Xi}^{-1}\preceq  ({\bm \Xi}-{\bm \Theta})^{-1}$ and $[{\bm \Xi}^{-1}]_{ii}\leq [({\bm \Xi}-{\bm \Theta})^{-1}]_{ii}$, $\forall{i}$, if and only if ${\bm \Theta}$ is a zero matrix. \hfill$\blacksquare$

\textbf{Theorem 1}. \emph{An approximate (loose) CRLB of the joint location and velocity estimation of a single target is expressed by 
\begin{align}
{\bm C}_{\rm app} = ({\bm \aleph}{\bm \varPhi} {\bm \aleph}^{\rm T})^{-1},
\end{align} 
which satisfies
${\bm C}_{\rm app}\preceq  {\bm C}_{\rm LV}$,  and $[{\bm C}_{\rm app}]_{ii}\leq [{\bm C}_{\rm LV}]_{ii}$, $\forall{i}$. The tightness of the approximate CRLB can be achieved, i.e., ${\bm C}_{\rm app}={\bm C}_{\rm LV}$, if and only if ${\bm \varPsi}$ is a zero matrix.}

\emph{Proof.}  The original CRLB is given in \eqref{eq:CRLB LV}.
According to the derivations above, we know that ${\bm \aleph}{\bm \varPhi}{\bm \aleph}^{\rm T}$ is positive definite symmetric matrix, and ${\bm \aleph}{\bm \varPsi}{\bm F}^{-1}{\bm \varPsi}^{\rm T}{\bm \aleph}^{\rm T}$ is positive semi-definite symmetric matrix. They satisfy ${\bm \aleph}{\bm \varPhi} {\bm \aleph}^{\rm T}\succ {\bm \aleph}{\bm \varPsi}{\bm F}^{-1}{\bm \varPsi}^{\rm T}{\bm \aleph}^{\rm T}$. According to \textbf{Lemma 1}, we have ${\bm C}_{\rm app}\preceq {\bm C}_{\rm LV}$  and $[{\bm C}_{\rm app}]_{ii}\leq [{\bm C}_{\rm LV}]_{ii}$, $\forall{i}$, where both equalities hold, if and only if ${\bm \varPsi}$ is a zero matrix. \hfill$\blacksquare$

\textbf{Theorem 2}. \emph{A sufficient condition to ensure the accuracy (tightness) of the approximate CRLBs is to employ a sensing waveform with a relatively large bandwidth and pulse time width, such that $\bar{f}_n^2\ll \bar{f_n^2}$ and $(\bar{t}_n+\tau_{n,k})^2\ll \bar{t_n^2}$}, $\forall{n, k}$. 

\emph{Proof.} The detailed proof of is given in Appendix \ref{sec:CRLB approx}. \hfill$\blacksquare$

\subsubsection{Multiple-Target CRLB}\label{sec:MT CRLB}
We now derive the CRLB of the joint location and velocity estimation for multiple-target sensing.
Firstly, we calculate the partial derivative matrix 
   $\frac{\partial {\bm \phi}_{\rm MT}}{\partial {\bm \theta}_{\rm MT}} = {\rm diag}\{{\bm \varLambda}_1,{\bm \varLambda}_2,\cdots,{\bm \varLambda}_Q\}$,
where ${\bm \varLambda}_q=\frac{\partial {\bm \phi}_q}{\partial {\bm \theta}_q}={\rm diag}\{{\bm \aleph}_q,{\bm I}_{2NK}\}$ with target-wise parameter ${\bm \theta}_q$ and ${\bm \phi}_q$ defined in Section \ref{sec:discrete LLF} and ${\bm \aleph}_q$ defined similarly as \eqref{eq:FO-phi-theta}. The FIM ${\bm J}_{\rm MT}({\bm \phi}_{\rm MT})$ is given by 
\begin{equation}
   \small
   {\bm J}_{\rm MT}({\bm \phi}_{\rm MT})=\begin{bmatrix}
      {\bm J}_{11}&{\bm J}_{12}&\cdots&{\bm J}_{1Q}\\{\bm J}_{21}&{\bm J}_{22}&\cdots&{\bm J}_{2Q}\\\cdots&\cdots&\cdots&\cdots\\{\bm J}_{Q1}&{\bm J}_{Q2}&\cdots&{\bm J}_{QQ}
   \end{bmatrix},\label{eq:FIM phi multiple}
\end{equation}
where for $1\leq q, l\leq Q$, and 
\begin{equation}
   \small
   {\bm J}_{ql}\triangleq -\mathbb{E}_{{\bm r}}\left\{\frac{\partial^2 \log{p({{\bm r}}|{\bm \phi})}}{\partial{\bm \phi}_q{\partial{\bm \phi}_l^{\rm T}}}\right\}=\begin{bmatrix}
   {\bm A}_{ql}&{\bm B}_{ql}&{\bm G}_{ql}\\\tilde{\bm B}_{ql} &{\bm D}_{ql}&{\bm E}_{ql}\\\tilde{\bm G}_{ql} &\tilde{\bm E}_{ql} &{\bm F}_{ql}
   \end{bmatrix}.\label{eq:FIM_phi q}
\end{equation}
In particular, we have $\tilde{\bm B}_{qq}={\bm B}_{qq}^{\rm T}$, $\tilde{\bm G}_{qq}={\bm G}_{qq}^{\rm T}$ and $\tilde{\bm E}_{qq}={\bm E}_{qq}^{\rm T}$ for $1\leq q\leq Q$.
The CRLB matrix for ${\bm \theta}$ is ${\bm C}_{\rm MT} = {\bm J}_{\rm MT}^{-1}({\bm \theta}_{\rm MT})$  with 
\begin{equation}
   \small
\begin{aligned}
    \setlength{\arraycolsep}{1pt}
   &{\bm J}_{\rm MT}({\bm \theta}_{\rm MT})  =\frac{\partial {\bm \phi}_{\rm MT}}{\partial{\bm \theta}_{\rm MT}}{\bm J}_{\rm MT}({\bm \phi}_{\rm MT})\left(\frac{\partial {\bm \phi}_{\rm MT}}{\partial{\bm \theta}_{\rm MT}}\right)^{\rm T}\\
   & ~~~~~~~ =\begin{bmatrix}
      {\bm \varLambda}_1{\bm J}_{11}{\bm \varLambda}_1&{\bm \varLambda}_1{\bm J}_{12}{\bm \varLambda}_2&\cdots&{\bm \varLambda}_1{\bm J}_{1Q}{\bm \varLambda}_Q\\{\bm \varLambda}_2{\bm J}_{21}{\bm \varLambda}_1&{\bm \varLambda}_2{\bm J}_{22}{\bm \varLambda}_2&\cdots&{\bm \varLambda}_2{\bm J}_{2Q}{\bm \varLambda}_Q\\\cdots&\cdots&\cdots&\cdots\\{\bm \varLambda}_Q{\bm J}_{Q1}{\bm \varLambda}_1&{\bm \varLambda}_Q{\bm J}_{Q2}{\bm \varLambda}_2&\cdots&{\bm \varLambda}_Q{\bm J}_{QQ}{\bm \varLambda}_Q
   \end{bmatrix}.\label{eq:FIM theta multiple}
   \end{aligned}
\end{equation}
The FIM block ${\bm J}_{ql}$ in \eqref{eq:FIM phi multiple} and \eqref{eq:FIM theta multiple} is calculated by taking the negative Hessian matrix of \eqref{eq:LF multiple} regarding the parameters of each target pair $\{q,l\}$.
Due to space constraints, we refer the reader to Appendix \ref{sec:CRLB multiple} for further details.

\subsection{CRLB Analysis}
In this section, we analyze and discuss the theoretical performance of DMS MIMO radar sensing and summarize the key findings.

We first summarize the impact of the number of samples (or sampling rate) and the SCNR on the CRLB for both single- and multiple-target sensing in a DMS MIMO radar.

\textbf{Theorem 3}. \emph{In both single-target and multiple-target scenario, the CRLB of the joint location and velocity estimation is inversely proportional to the number of samples (or equivalently, sampling rate) and the SCNR, i.e., ${\bm C}\propto \frac{1}{S\delta}$}.

\emph{Proof}. The product $S\delta$ is a multiplicative factor in corresponding FIMs ${\bm J}(\phi)$ of \eqref{eq:FIM_phi} and ${\bm J}_{ql}$ of \eqref{eq:FIM theta multiple}. Based the relationship between FIM and CRLB, the CRLB is inversely proportional to both the number of sampling $S$ and the SCNR $\delta$. \hfill$\blacksquare$

According to \textbf{Theorem 3}, the adverse effects of low SCNR can be alleviated by increasing the number of samples. However, increasing the sample count results in higher computational complexity and energy consumption, while achieving a larger SCNR leads to increased power expenditure. Consequently, a trade-off must be carefully evaluated among computational complexity, power budget, and localization accuracy to ensure an efficient and balanced system design.

Based on the derived approximate CRLBs for single target sensing, \textbf{Theorem 4} further investigates the impact of key parameters, namely, the number of samples, SCNR and SEBW, on the CRLB for single static target localization, i.e, $v=0$. 

\textbf{Theorem 4}. \emph{The CRLB for single static target localization is inversely proportional to the number of samples, SCNR and the SEBW, i.e., $C_{\rm static}\propto \frac{1}{S\delta\bar{f_n^2}}$.}

\emph{Proof}. The approximated CRLB for static target localization can be straightforward from the result in \textbf{Theorem 1} by removing the Doppler-frequency-related terms, i.e.,
\begin{align}
   C_{\rm static} = ({\bm \aleph}_{11}{\bm A}{\bm \aleph}_{11}^{\rm T})^{-1}, \label{eq:CRLB loc Approx}
\end{align}
where ${\bm \aleph}_{11}$ characterizes the geometric spread of the DMS MIMO radar, and ${\bm A}$ denotes the FIM regarding delay. The $((n-1)K+k)$th diagonal element of ${\bm A}$ is given by 
\begin{align}
[{\bm A}]_{(n-1)K+k,(n-1)K+k}=a_{n,k}=8\pi^2|\alpha_{n,k}|^2\rho_{n}S\delta\bar{f_n^2}
\end{align}
Consequently, the single-target localization CRLB is inversely proportional to the number of samples $S$, SCNR $\delta$ and the SEBW $\bar{f_n^2}$. \hfill$\blacksquare$

When considering joint location and velocity estimation, the relationship between the CRLBs and SEBW (or SETW) is not straightforward. 

\textbf{Theorem 5}. \emph{With a band-limited Gaussian waveform, the (approximate) location CRLB is inversely proportional to the SETW, i.e., ${\bm C}_{\rm L}\propto \frac{1}{\bar{t_n^2}}$, but insensitive to the SEBW ${\bar{f_n^2}}$.}

\emph{Proof.} The proof is given in Appendix \ref{sec: proof of theorem 5}. \hfill$\blacksquare$

Note that SEBW can be adjusted independently without altering the SETW, for example, by varying the number of subcarriers in the OCDM signal in \eqref{eq:OCDM}. In this case, as stated in \textbf{Theorem 5}, the location CRLB remains nearly unaffected.

In a multiple-target sensing scenario, multiple-target (coupling) interference will influence the CRLB of each target compared to that of a single-target sensing scenario. An approximate multiple-target CRLB is proposed by neglecting the direct structural coupling between target pairs. Based on that, we have analyzed the relationship between the single-target CRLB and the multiple-target CRLB in Appendix \ref{sec:CRLB trend}. 

\textbf{Theorem 6}. \emph{The sufficient condition for the multiple-target CRLB approaching the corresponding approximate or single-target CRLB is the large separation of different targets in delay or Doppler domain, i.e., $\forall{n,k,q,l}$,
\begin{align}
|\tau_{n,k,q,l}|\geq \tau_{\rm r}, \label{eq:delay resolution}
\end{align}
or
\begin{align}
   |f_{n,k,q,l}|\geq f_{\rm r}. \label{eq:Doppler resolution}
\end{align}}

\emph{Proof.} The proof is given in Appendix \ref{sec:CRLB trend}. \hfill$\blacksquare$

For any two targets, we define the safety distance as the minimum distance where two targets can be seen as independent targets, expressed by $d_r=\tau_{\rm r}c/2$ based on \eqref{eq:delay resolution}. Given $\tau_{\rm r} \propto  1/f_{\rm r}$, we further have 
\begin{align}
   d_r\propto  \frac{c}{2f_{\rm r}}.
\end{align}
Therefore, the safety distance is inversely proportional to the bandwidth. 

In particular, for two targets in the same location ${\bm l}_q$, their relative Doppler frequency can be expressed as 
\begin{align}
   f_{n,k,q,l} = \dfrac{1}{\lambda}{\bm v}_{q,l}^{\rm T}{\bm l}_{n,k,q}, \label{eq:relative Doppler}
\end{align}
where ${\bm l}_{n,k,q} = \frac{{\bm l}_n-{\bm l}_q}{\|{\bm l}_n-{\bm l}_q\|_2}+\frac{{\bm l}_k-{\bm l}_q}{\|{\bm l}_k-{\bm l}_q\|_2}$ is the location vector of target $q$ relative to transmitter $n$ and receiver $k$, and ${\bm v}_{q,l} = {\bm v}_{q}-{\bm v}_{l}$ denotes the relative velocity. Define the relative radial velocity as $\bar{v}_{q,l} = {\bm v}_{q,l}^{\rm T}{\bm l}_{n,k,q}/\|{\bm l}_{n,k,q}\|_2$. According to Cauchy-Schwarz inequality \cite{Mitrinovic1993} and triangle inequality \cite{Mitrinovic1993Tri}, we have $|\bar{v}_{q,l}|\leq \|{\bm v}_{q,l}\|_2$ and $\|{\bm l}_{n,k,q}\|_2\leq 2$, respectively. Consequently, \eqref{eq:relative Doppler} satisfies 
\begin{align}
   |f_{n,k,q,l}|&  {=} \dfrac{1}{\lambda}|\bar{v}_{q,l}|\|{\bm l}_{n,k,q} \|_2  \overset{(\text{i})}{\leq}  \dfrac{2|\bar{v}_{q,l}|}{\lambda}\overset{(\text{ii})}{\leq} \dfrac{2\|{\bm v}_{q,l}\|_2}{\lambda}, \label{eq:modulus rel Dopp}
\end{align}
where equality $(\mathrm i)$ holds when using a mono-static transceiver with $n=k$, and equality $(\mathrm {ii})$ holds when ${\bm v}_{q,l}$ and ${\bm l}_{n,k,q}$ have the same or opposite directions. 

Consequently, according to \eqref{eq:Doppler resolution} and \eqref{eq:modulus rel Dopp}, one has the safety velocity as
\begin{align}
   v_r \geq \frac{\lambda f_{\rm r}}{2}\propto \frac{\lambda}{2\tau_{\rm r}}.
\end{align}
In particular, when using OFDM subcarrier for mono-static radar, we have $T_{\rm eff}=\tau_{\rm r}=1/f_{\rm r}=1/B_{\rm eff}$, and thus $d_r= \frac{c}{2B_{\rm eff}}$ and $v_r = \frac{\lambda}{2T_{\rm eff}}$.

\section{Simulation results}\label{sec:simulations}
In this section, we conduct simulations to evaluate the derived CRLBs under two scenarios: (i) pure localization, and (ii) joint location and velocity estimation. All access points (APs), comprising $N$ single-antenna transmitters and $K$ single-antenna receivers, are assumed to be stationary.
To ensure consistency with prior studies \cite{Godrich2010, Godrich2011, Godrich2012, Ai2015, he2012noncoherent, He2010, Chuanming2010}, we adopt the signal energy-to-noise ratio (SENR), defined as $\tilde\delta\triangleq P/\sigma_z^2$, instead of the SCNR $\delta$ in evaluating the CRLBs. 

\begin{figure}[!t] 
   \centering
   \subfigure[]{\label{fig:Location Radar Symm}\includegraphics[width=0.22\textwidth]{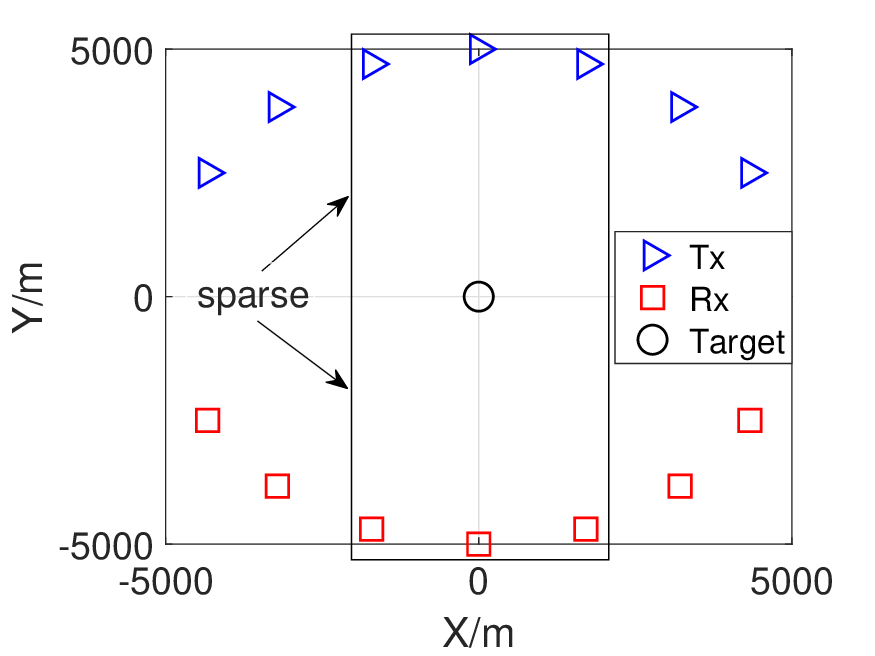}}
   \subfigure[]{\label{fig:Location Radar Non}\includegraphics[width=0.22\textwidth]{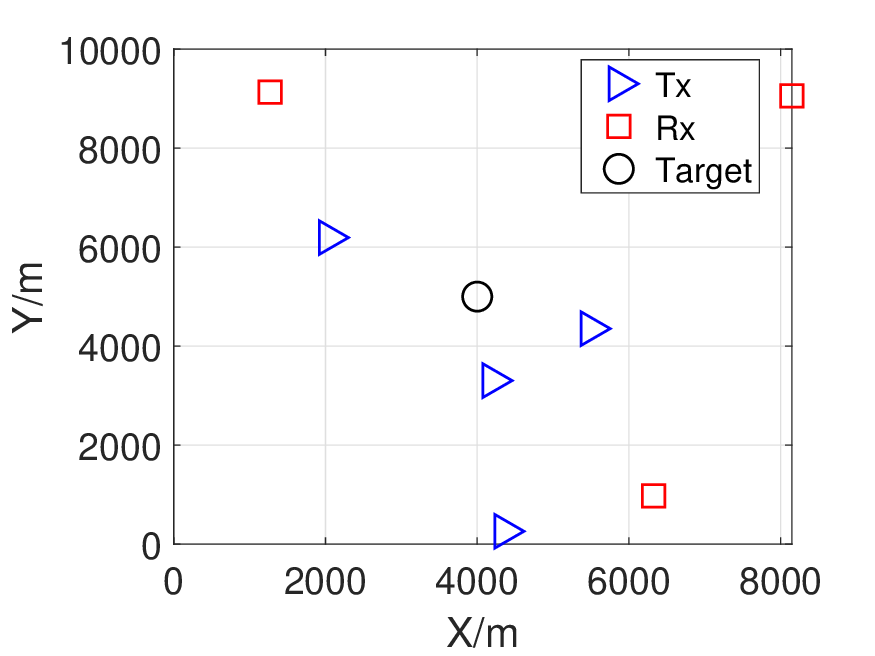}}
   \caption{The location setting of the MIMO radar system: (a) $7\times 7$ symmetric setting; (b) $4\times 3$ nonsymmetric setting.}
   \label{fig:Location Radar}
   \vspace{-0.3cm}
\end{figure}

Two radar configurations are employed in this section. As illustrated in Fig. \ref{fig:Location Radar Symm}, a circularly symmetric multi-static radar configuration, with transmitters located at azimuths $\{30^\circ,50^\circ,70^\circ,90^\circ,110^\circ,130^\circ,150^\circ\}$ and receivers positioned at azimuths $\{-30^\circ,-50^\circ,-70^\circ,-90^\circ,-110^\circ,-130^\circ,-150^\circ\}$ by a radius of $R=5$ km. Fig. \ref{fig:Location Radar Non} demonstrates a $4\times 3$ asymmetric setting. Unless noted otherwise, the symmetric setting in Fig. \ref{fig:Location Radar Symm} is considered.
\subsection{Localization CRLB}\label{sec:pure localization sim}
Two basic waveforms are considered for localization: OCDM \cite{Ouyang2016} and OFDM. A single Gaussian pulse shaping is applied to both waveforms. We consider Frequency Range 1 \cite{Nuria2024}, with a carrier frequency of 3 GHz for both waveforms. Mutually orthogonal sensing subcarriers are used across different transmitters, $n=1,2,\cdots,N$. The lowpass equivalents of the subcarriers for OCDM and OFDM waveforms are respectively expressed as 
\begin{align}
s_n(t) = (2/T^2)^{1/4}e^{-\pi t^2/T^2}e^{j\pi M/T^2(t-(n-1)T/M)^2}, \label{eq:OCDM}
\end{align}
and
\begin{align}
   s_n(t) = (2/T^2)^{1/4}e^{-\pi t^2/T^2}e^{j2\pi (n-1)t/T},
\end{align}
where $M$ denotes the available number of OCDM chirp subcarriers, and $T$ is proportional to the effective pulse time width. Note that the bandwidth of OCDM subcarrier is about $M$ times larger than that of the OFDM subcarrier.  
\begin{figure}[!t]
   \centerline
   {\includegraphics[width=0.3\textwidth]{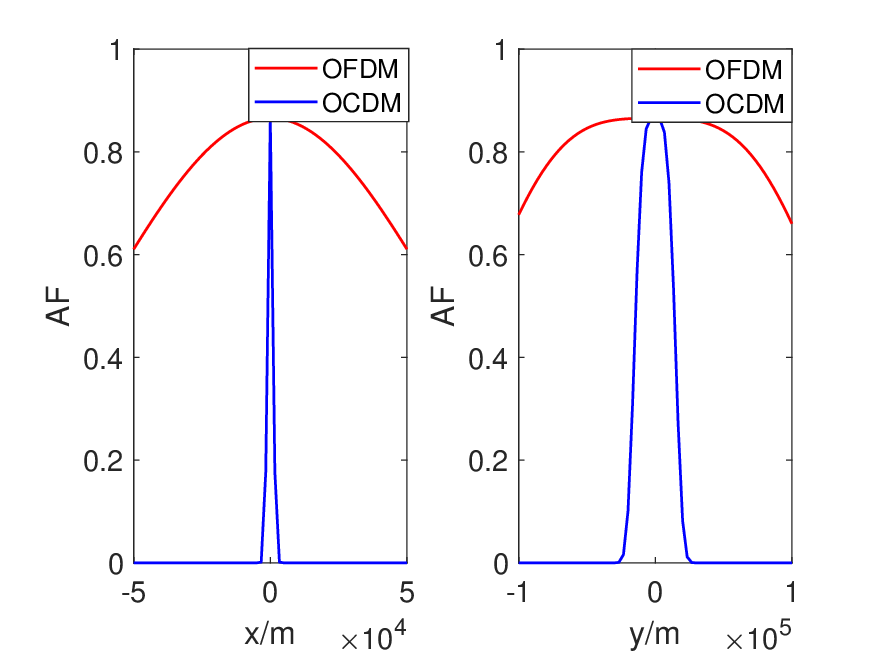}}
   \caption{The ambiguity function under different waveforms.}\label{fig:AF}
   \vspace{-0.3cm}
\end{figure}
\begin{figure}[!t]
   \centerline
   {\includegraphics[width=0.3\textwidth]{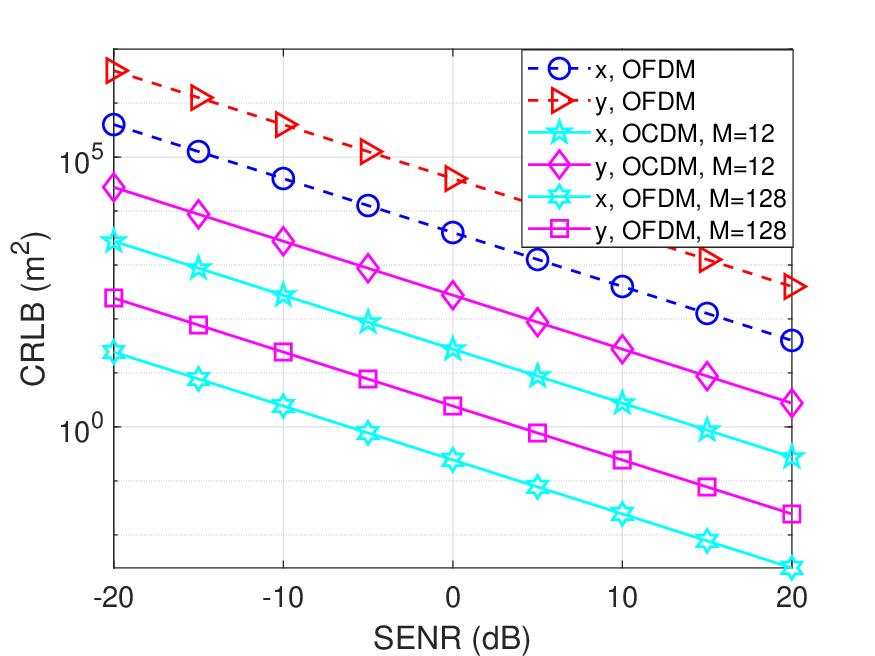}}
   \caption{The CRLB regarding the SENR.}\label{fig:CRLB Wave BW}
   \vspace{-0.3cm}
\end{figure}
\begin{figure}[!t]
   \centering
   \subfigure[]{\label{fig:CRLB MT local OFDM}\includegraphics[width=0.3\textwidth]{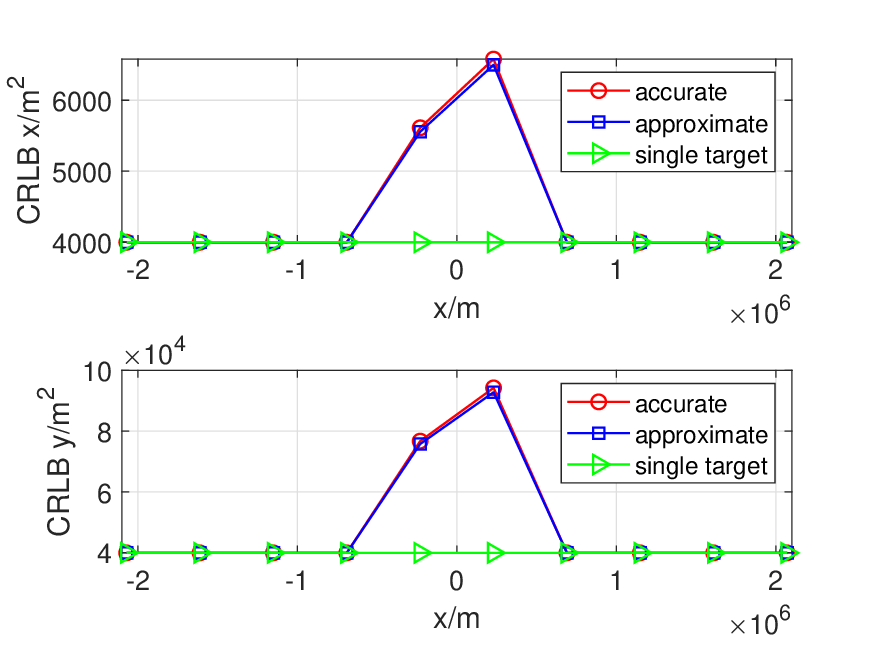}}
   \subfigure[]{\label{fig:CRLB MT local OCDM}\includegraphics[width=0.3\textwidth]{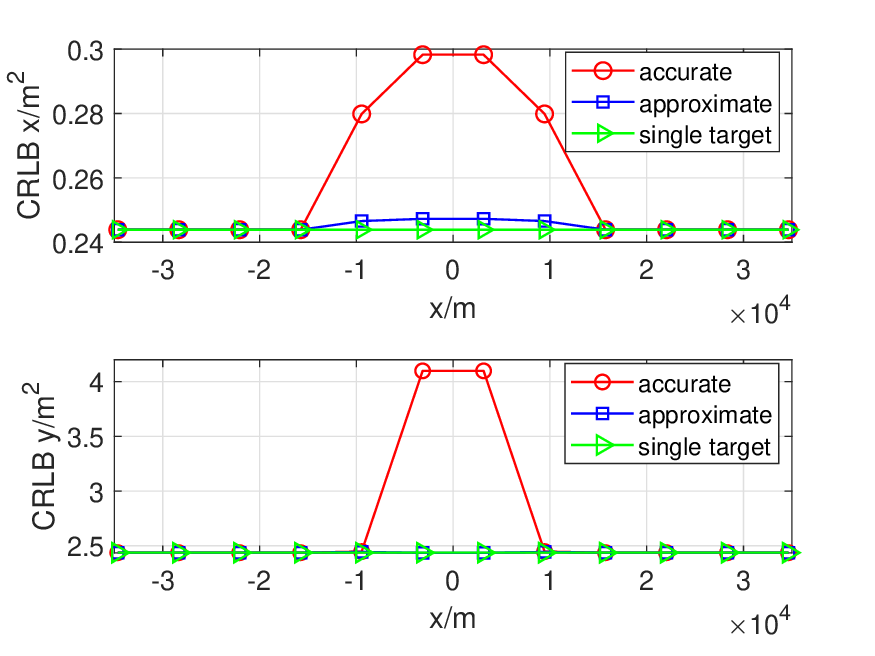}}
   \caption{The localization CRLB regarding the relative distance: (a) OFDM; (b) OCDM.}
   \label{fig:CRLB MT local}
   \vspace{-0.3cm}
\end{figure}

We begin by analyzing the ambiguity functions (AFs) as defined in \eqref{eq:AF discrete}. As an illustrative example, consider a colocated transceiver located at $(1 \times 10^5, 0)$ m and a single target positioned at the origin $(0,0)$ m. The corresponding AFs are shown in Fig.~\ref{fig:AF}. It is observed that the OCDM waveform produces a narrower AF beamwidth compared to the OFDM waveform, owing to its higher bandwidth. This narrower beamwidth indicates superior sensing resolution and accuracy with OCDM waveforms, which is also evidenced by \eqref{eq:CRLB loc Approx} in \textbf{Theorem 4}.

We now evaluate the CRLBs versus the SENR and the bandwidth. Without loss of generality, assume a single static target at the origin $(0,0)$ m with its RCS matrix given as ${\bm A}_1 = \sqrt{1/2}({\bm 1}_{N\times K}+j{\bm 1}_{N\times K})$. For both waveforms, assume $T=10^{-3}$ s and a sampling rate $f_s=100$ kHz. For the OCDM, two configurations of $M$ are considered: $M=12$ and $M=128$. The CRLBs regarding the SENRs are shown in Fig. \ref{fig:CRLB Wave BW}. The proportional relation is firstly observed between CRLB and the SENR (or SCNR) for both waveforms. We also find that significantly lower CRLBs are achieved when using the OCDM waveform, particularly with larger $M$. This is because, when $M>N$, the bandwidth of OCDM waveform is larger than the total bandwidth of $N$ OFDM subcarriers over $N$ transmitters. These results align with \textbf{Theorem 4}. 
 
We now evaluate the multiple-target localization CRLBs. Consider a second target located at $(x,0)$ m, with a smaller RCS matrix given by ${\bm A}_2=1/5{\bm A}_1$. The SENR is set to $0$ dB, and $M=128$ for the OCDM waveform. Fig.~\ref{fig:CRLB MT local} illustrates the CRLB for the first target as a function of the second target's relative distance along the x-axis, using both OFDM and OCDM waveforms.
The \emph{single-target} CRLB is derived in Section~\ref{sec:ST CRLB}, while the \emph{accurate} and \emph{approximate} multiple-target CRLBs are provided in Section~\ref{sec:MT CRLB} and Appendix~\ref{sec:CRLB trend}, respectively. As the relative distance between the two targets increases (in either direction), the interference from the second target becomes negligible, and the CRLB for the first target converges toward that of the single-target case.
Additionally, due to its higher bandwidth, the OCDM waveform yields better resolution (smaller safety distance, x-axis) and improved localization accuracy (i.e., smaller CRLBs along the y-axis) compared to OFDM. These trends are evident in Figs.~\ref{fig:CRLB MT local OFDM} and \ref{fig:CRLB MT local OCDM}, and are consistent with the theoretical insights presented in \textbf{Theorem 4} and \textbf{Theorem 6}. Hence, OCDM is adopted for use in the subsequent analysis.
\subsection{Joint location and velocity estimation} 
We now evaluate the CRLBs for joint location and velocity estimation in both single-target and multiple-target sensing scenarios. Specifically, we conduct the simulation from the following aspects: 1) Deep fading and large-scale MIMO radar;
2) Receiver sampling rate;
3) The proposed approximate CRLB for location and velocity;
4) Multiple-target CRLB;
5) Comparison between the MLE and the derived CRLBs.
\begin{figure}[!t]
   \centering
   \subfigure[]{\label{fig:CRLB L OCDM DF}\includegraphics[width=0.3\textwidth]{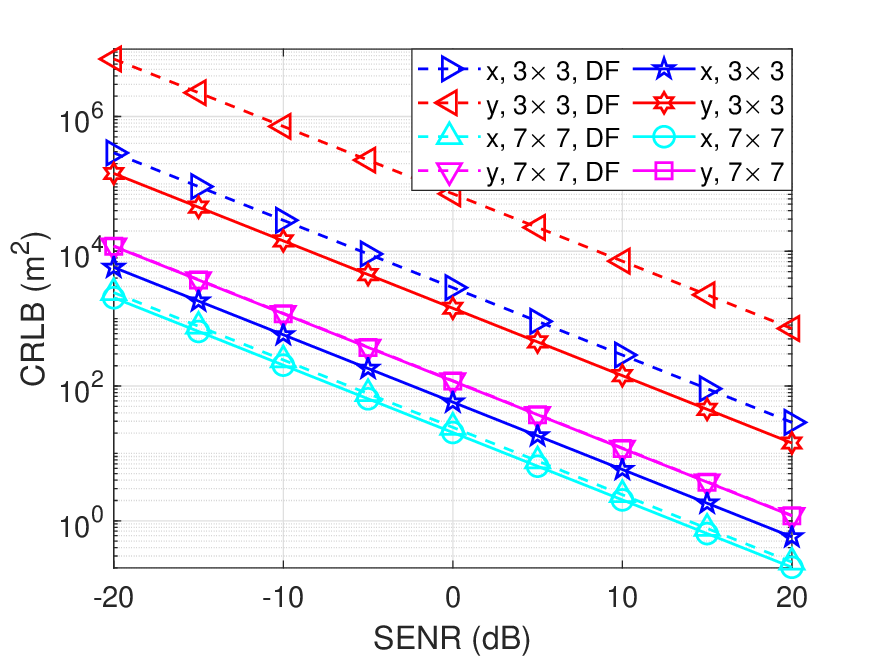}}
   \subfigure[]{\label{fig:CRLB V OCDM DF}\includegraphics[width=0.3\textwidth]{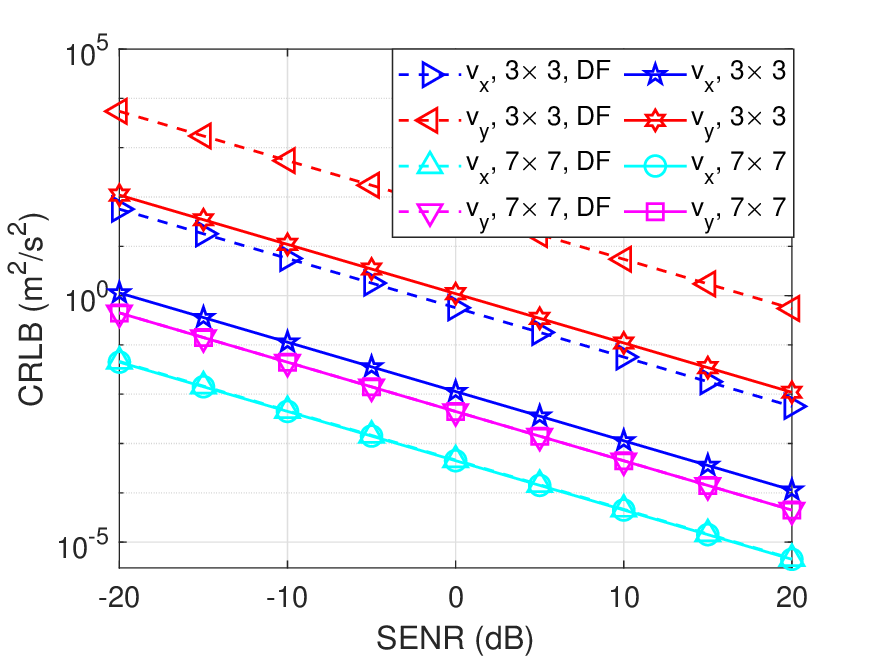}}
   \caption{The multi-static CRLBs under deep fading (DF): (a) location; (b) velocity.}
   \label{fig:CRLB OCDM DF}
   \vspace{-0.3cm}
\end{figure}
\begin{figure}[!t]
   \centerline
   {\includegraphics[width=0.3\textwidth]{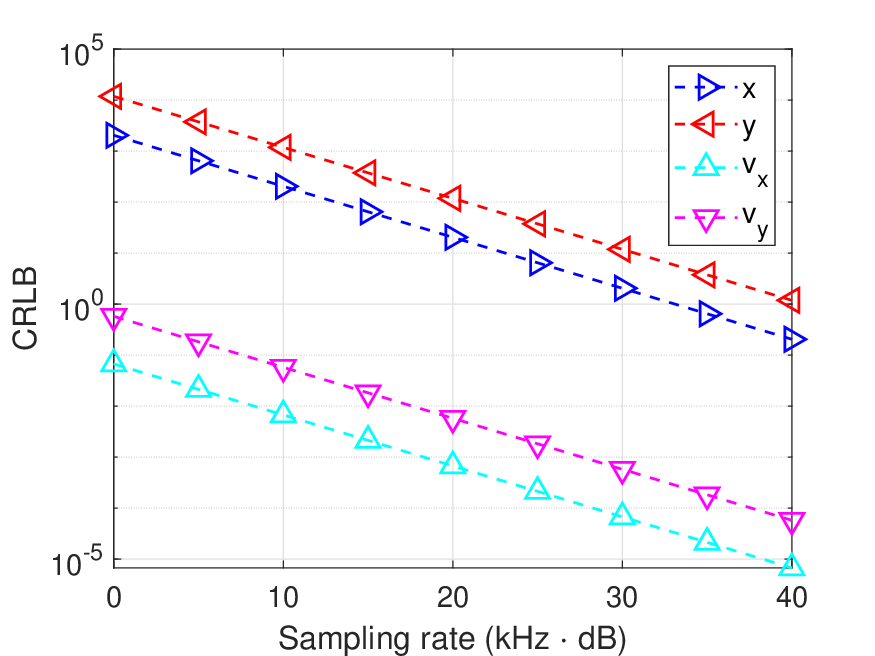}}
   \caption{The CRLB regarding the sampling rate.}\label{fig:CRLB fs}
   \vspace{-0.3cm}
\end{figure}

We consider a single target located at the origin $(0,0)$ m, moving with a velocity of $(-15, 0)$ m/s. Assume uniform RCS values of $\sqrt{1/2}(1 + j)$ for all links in Fig. \ref{fig:Location Radar Symm}. 
Firstly, to evaluate the effect of the number of antennas on the CRLB, a sparser radar configuration is examined, comprising three transmitters placed at azimuths $\{70^\circ, 90^\circ, 110^\circ\}$ and three receivers at $\{-70^\circ, -90^\circ, -110^\circ\}$, as indicated by the rectangular area in Fig. \ref{fig:Location Radar Symm}. 
Secondly, to evaluate the influence of deep fading (DF), the $3\times 3$ sparse radar setting is assumed to be deep faded (DF) with lower RCSs ${\bm A}_s=0.1({\bm 1}_{2\times 2}+j{\bm 1}_{2\times 2})$ while the remaining RCSs are unchanged. The OCDM waveform is used in two cases, with parameters $M = 128$, $T = 10^{-2}$ s, and a sampling rate $f_s = 1$ kHz. 

Simulation results are presented in Fig.~\ref{fig:CRLB OCDM DF}, with each configuration labeled according to the radar setting and fading condition. As expected and consistent with \textbf{Theorem 3}, the CRLBs are observed to be inversely proportional to the signal-to-energy-plus-noise ratio (SENR) or SCNR. Moreover, deep fading leads to a significant increase in CRLBs due to the degraded reception SCNR. However, employing a large-scale multi-static radar network effectively mitigates this degradation through enhanced spatial diversity. Additionally, increasing the number of transceivers improves overall performance even in non-fading conditions, aligning with the findings reported in \cite{He2010}.

Following the same $7 \times 7$ symmetric system configuration without deep fading as in Fig. \ref{fig:CRLB OCDM DF}, Fig. \ref{fig:CRLB fs} shows the CRLB versus the sampling rate at an SENR of $-20$ dB. To represent proportional changes in the sampling rate with equal intervals on the x-axis, we use the unit ${\rm kHz}\cdot {\rm dB}$ with $f_1 {\rm kHz}\cdot {\rm dB}=10^{f_1/10}{\rm kHz}$.
We observe that CRLB improves proportionally with the increase of the sampling rate. Furthermore, compared to Fig. \ref{fig:CRLB OCDM DF}, it becomes evident that increasing the sampling rate yields a similar performance enhancement as increasing the SENR. This observation underscores a fundamental trade-off between computation and energy resources in sensing system design, as discussed in \textbf{Theorem 3}.

\begin{figure}[!t]
   \centering
   \subfigure[]{\label{fig:CRLB L OCDM APP}\includegraphics[width=0.3\textwidth]{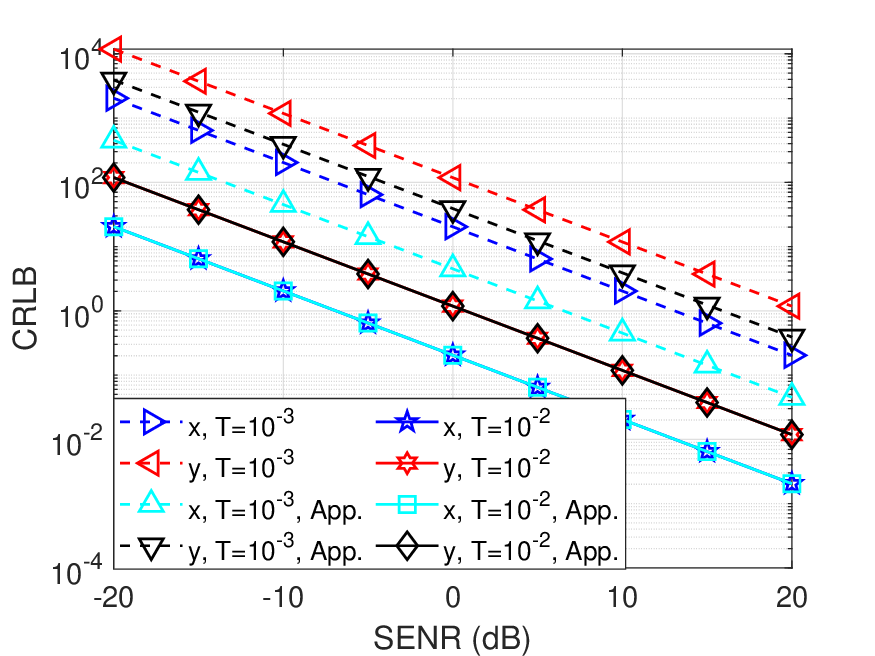}}
   \subfigure[]{\label{fig:CRLB V OCDM APP}\includegraphics[width=0.3\textwidth]{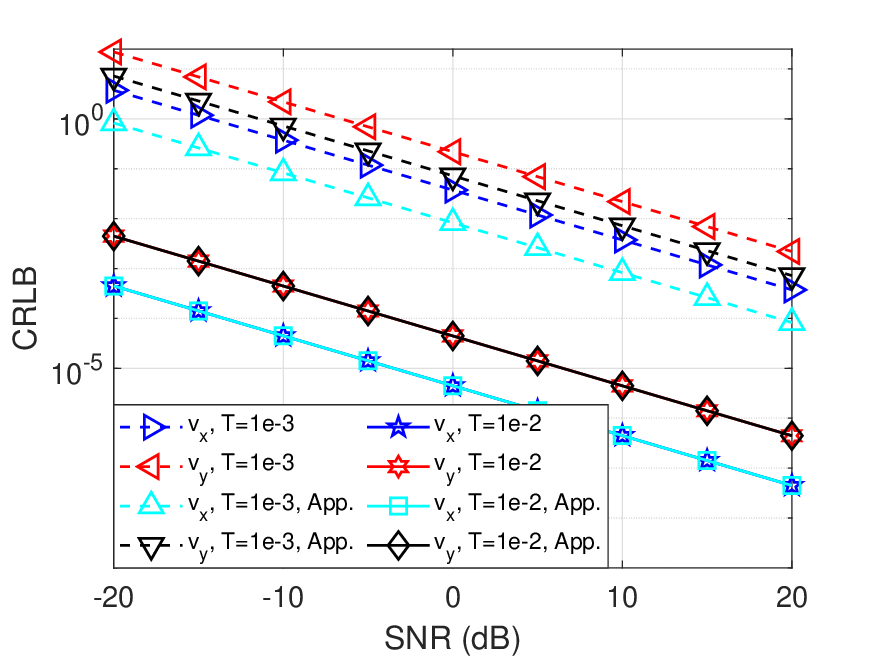}}
   \caption{The approximate CRLBs regarding the SENR: (a) location; (b) velocity.}
   \label{fig:CRLB OCDM APP}
   \vspace{-0.3cm}
\end{figure}

Fig.~\ref{fig:CRLB OCDM APP} illustrates the approximate CRLBs under two SETWs, corresponding to $T = 10^{-2}$ s and $T = 10^{-3}$ s. The bandwidth parameter and sampling rate are fixed at $M = 128$ and $f_s = 100$ kHz, respectively. Based on the OCDM waveform definition in \eqref{eq:OCDM}, we have $\bar{t}_n^2 \propto T$ when other parameters are held constant. As shown in the figure, the approximate CRLBs (denoted as “App.”) exhibit excellent agreement with the accurate CRLBs when $T$ is large, but become increasingly loose for smaller $T$, due to the reduced SETW. This observation validates the sufficient condition for approximation accuracy established in \textbf{Theorem 2}. Note that no comparison is made in terms of the bandwidth variation, as the bandwidth is sufficiently large to satisfy the approximation condition. Furthermore, we observe that both the location and the velocity estimation performance improve with increasing SETW, consistent with the trend predicted by \textbf{Theorem 5}..

\begin{figure}[!t]
   \centerline
   {\includegraphics[width=0.3\textwidth]{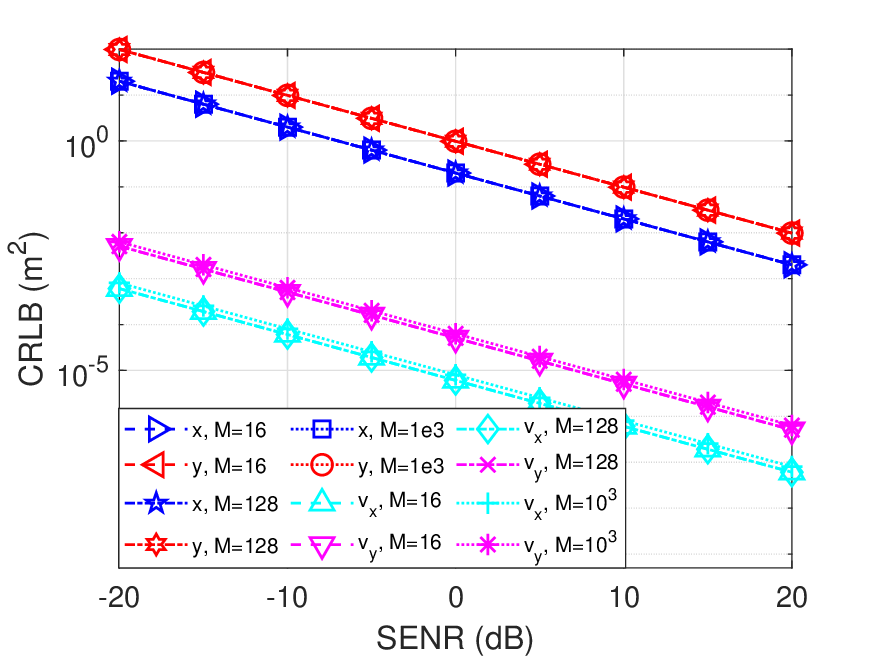}}
   \caption{The CRLBs regarding the SENR under different bandwidths.}\label{fig:CRLB OCDM BW}
   \vspace{-0.3cm}
\end{figure}

Furthermore, given $T=10^{-3}$ s and $f_s=100$ kHz, the effect of the SEBW on the CRLBs is examined in Fig.~\ref{fig:CRLB OCDM BW} by varying $M$. It is observed that the location CRLBs remain nearly invariant across different SEBW values, while the velocity CRLBs also exhibit limited sensitivity to SEBW changes. These findings are consistent with the theoretical insights provided in \textbf{Theorem 5}.

\begin{figure}[!t]
   \centerline
   {\includegraphics[width=0.3\textwidth]{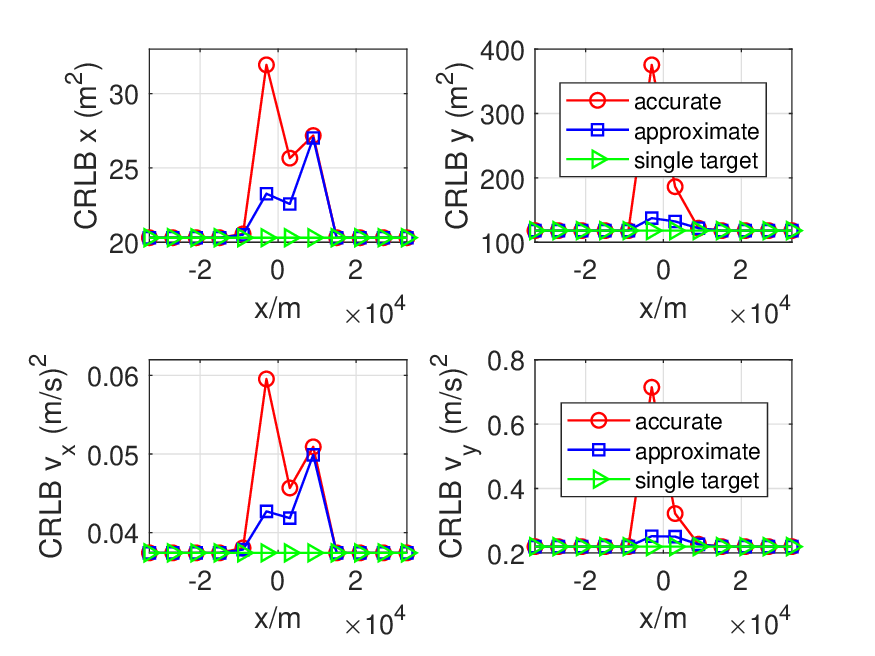}}
   \caption{The CRLB regarding the relative distance.}\label{fig:CRLB MT delay}
   \vspace{-0.3cm}
\end{figure}

Next, the impact of multi-target interference is examined under different spatial and velocity separations. We assume $M=128$, $T=10^{-3}$ s and $f_s=100$ kHz. The RCS matrix of the Target-$1$ is ${\bm A}_1 = \sqrt{1/2}({\bm 1}_{N\times K}+j{\bm 1}_{N\times K})$. Target-$2$ is separated from the Target-$1$ in time delay (distance) or Doppler frequency (velocity) with a RCS matrix ${\bm A}_2=1/5{\bm A}_1$. Assume Target-$2$ has the same velocity as Target-$1$ but distinguished location $(x,0)$ m for $x\neq 0$. Fig. \ref{fig:CRLB MT delay} shows the CRLB of Target-$1$ regarding Target-$2$'s relative distance along the x-axis, where SENR=$10$ dB is considered. Similar with the localization scenario in Section \ref{sec:pure localization sim}, when two targets are sufficiently spatially separated, the impact between each other can be neglected, and thus both the location and velocity CRLBs gradually approach their respective counterparts of the single-target sensing scenario. Furthermore, by comparing Fig. \ref{fig:CRLB MT delay} to Fig. \ref{fig:CRLB MT local OCDM}, we can observe a similar safety distance (or range resolution) for both the pure localization and the joint location and velocity estimation. This is because both scenarios employ the same bandwidth parameter.

\begin{figure}[!t]
   \centerline
   {\includegraphics[width=0.3\textwidth]{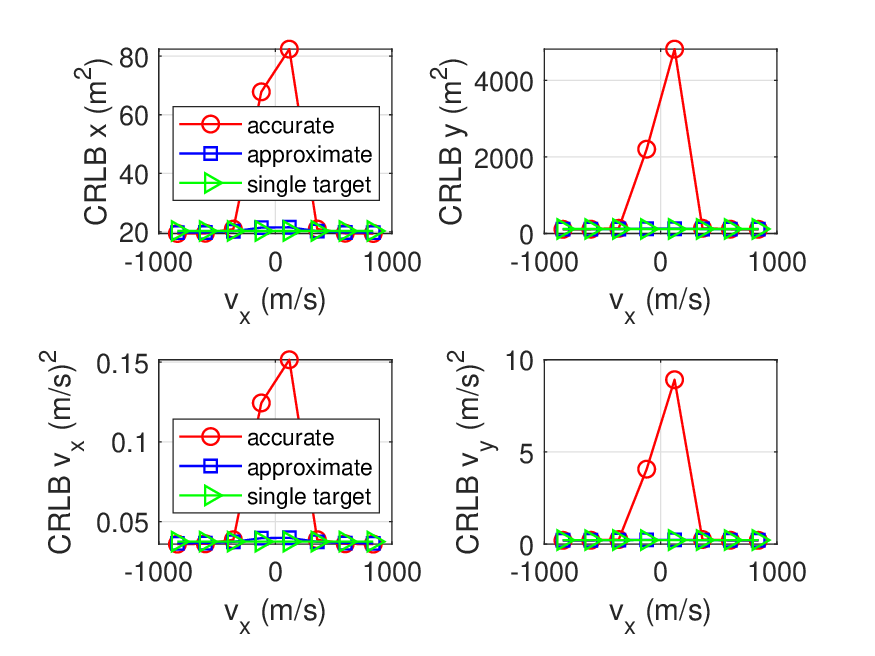}}
   \caption{The CRLB regarding the relative velocity.}\label{fig:CRLB MT Doppler}
   \vspace{-0.3cm}
\end{figure}

Assuming that Target-$2$ shares the same location as Target-$1$ but possesses a distinct velocity of $(v_x - 15, 0)$ m/s, Fig. \ref{fig:CRLB MT Doppler} illustrates the velocity CRLB of the Target-$1$ with respect to its relative velocity to Target-$2$. It is observed that when the two targets are sufficiently separated in the Doppler domain, namely, when the value of $\vert v_x \vert$ is sufficiently large, their CRLBs converge to their respective single-target counterparts. Furthermore, as observed from Figs. \ref{fig:CRLB MT local}, \ref{fig:CRLB MT delay} and \ref{fig:CRLB MT Doppler}, the approximate multiple-target CRLB exhibits smaller deviations from the single-target CRLB compared to the accurate CRLBs. This is because the approximate CRLBs disregard the direct structural coupling present in the off-diagonal FIM blocks ${\bm J}_{ql}$ and only retain the additive coupling within each diagonal block ${\bm J}_{qq}$, as analyzed in Appendix \ref{sec:CRLB trend}. 


\begin{figure}[!t]
   \centerline
   {\includegraphics[width=0.35\textwidth]{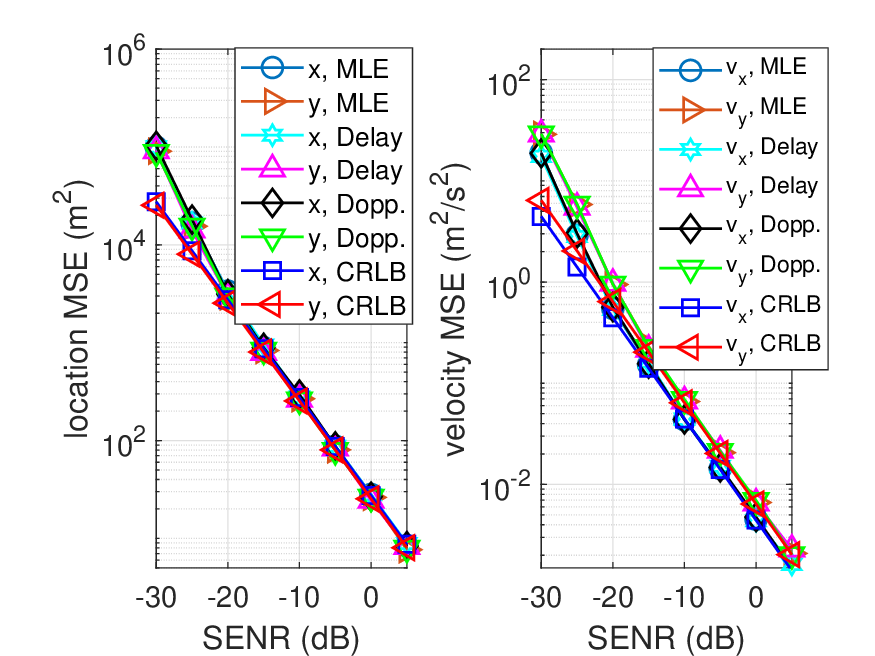}}
   \caption{The MSE versus the CRLB under different SENRs.}\label{fig:MSE CRLB}
   \vspace{-0.3cm}
\end{figure}

\begin{figure}[!t]
   \centerline
   {\includegraphics[width=0.3\textwidth]{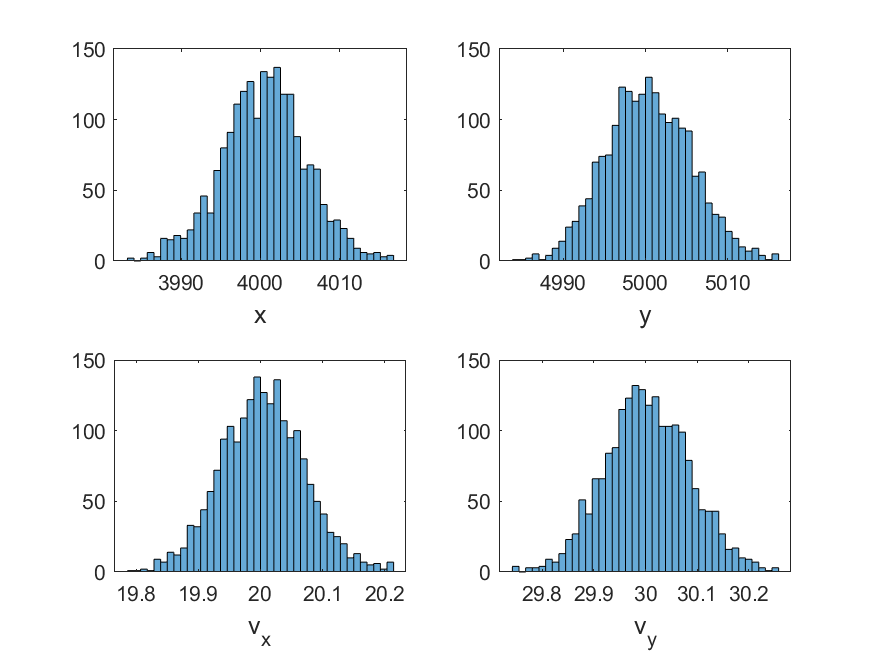}}
   \caption{The histogram for MLE at 0 dB SENR.}\label{fig:MLE hist}
   \vspace{-0.3cm}
\end{figure}

We now verify the CRLB in comparison with the MLE results mentioned in Section \ref{sec:ST MLE}. Due to the high complexity of four-dimensional MLE, we consider a simple $4\times 3$ nonsymmetric multi-static radar configuration given in Fig. \ref{fig:Location Radar Non}. Consider the OCDM waveform with $M=16$, $T=10^{-2}$ s and a sampling rate of $f_s=1$ kHz. The velocity of a single target is $(20,30)$ m/s. 
Fig. \ref{fig:MSE CRLB} compares the mean squared error (MSE) of the MLE with the derived CRLB. It is observed that the derived CRLBs closely match the MSEs for SENR values greater than $-20$ dB. The small discrepancies observed in the low SENR regime are attributed to the suboptimal behavior of the likelihood function in noise-dominated conditions.
Additionally, the histogram of location and velocity estimation is provided in Fig. \ref{fig:MLE hist}, suggesting that the MLE estimates follow a normal distribution under additive Gaussian clutter and noise.


Furthermore, consider the presence of Target-$2$, which is well separated from Target-1 either in the time-delay (distance) or Doppler frequency (velocity) domain. In this case, the joint location and velocity of Target-$1$ can be estimated using the proposed approximate multiple-target MLE method given in \eqref{eq:MLE sub function}. The corresponding results are labeled as ${\rm Delay}$ and ${\rm Dopp.}$ in Fig.~\ref{fig:MSE CRLB}. It is observed that the MSEs for both the location and velocity estimates of Target-$1$ remain largely unaffected by the presence of Target-$2$, due to their significant separation in either domain. This outcome further validates the effectiveness of the proposed MLE method \eqref{eq:MLE sub function} for accurately estimating parameters of targets that are sufficiently separated in delay or Doppler.
\section{Conclusions}\label{sec:Conclusions}
This paper has presented a comprehensive analysis of sensing performance in CF MIMO-ISAC systems, with a focus on joint location and velocity estimation under a deterministic and spatially diverse RCS assumption. For the first time, we derived CRLBs for both single-target and multiple-target scenarios using a discrete signal model that incorporates the impact of sampling rates. To address computational complexity, we developed closed-form approximate CRLBs and established sufficient conditions for their accuracy. Then, we formulated and discussed several theorems that characterize system performance based on the proposed (approximate) CRLB framework.
We also proposed the concepts of safety distance and safety velocity to characterize conditions under which multi-target CRLBs become effectively decoupled. Theoretical results and proposed CRLB formulations were validated through simulations using OFDM and OCDM waveforms, demonstrating strong consistency with MLE-based performance.

This work lays a foundation for practical sensing metric design in CF MIMO-ISAC systems. As a future direction, we will explore resource allocation strategies guided by the derived CRLBs to optimize joint communication and sensing performance.



\appendices
\normalsize

\section{Proof of \textbf{Theorem 2}}\label{sec:CRLB approx}

As demonstrated in \textbf{Lemma 1} and \textbf{Theorem 1}, to ensure the tightness of the approximate CRLBs, the second term inside the brackets of \eqref{eq:CRLB LV} must be negligible compared to the first term. This requirement is intuitive, as the matrix ${\bm \varPhi}$ captures both the first- and second-order delay and Doppler information, whereas ${\bm \varPsi}$ only includes first-order moment information. To facilitate analysis, we rewrite the CRLB expression from \eqref{eq:CRLB LV} as follows
\begin{align}
   {\bm C}_{\rm LV} = 
   ({\bm \aleph}({\bm \varPhi}-{\bm \varPsi}{\bm F}^{-1}{\bm \varPsi}^{\rm T}) {\bm \aleph}^{\rm T}
      )^{-1}. \label{eq:CRLB LV simplified}
\end{align}

It is easy to find that both ${\bm \varPhi}$ and ${\bm \varPsi}{\bm F}^{-1}{\bm \varPsi}^{\rm T}$ are positive semi-definite symmetric matrices. ${\bm \varPsi}{\bm F}^{-1}{\bm \varPsi}^{\rm T}$ can be interpreted as a perturbation to ${\bm \varPhi}$, rewritten as
\begin{align}
   {\bm \varPsi}{\bm F}^{-1}{\bm \varPsi}^{\rm T} = 
   \begin{bmatrix}{\bm G}{\bm F}^{-1}{\bm G}^{\rm T}&{\bm G}{\bm F}^{-1}{\bm E}^{\rm T}\\{\bm E}{\bm F}^{-1}{\bm G}^{\rm T}&{\bm E}{\bm F}^{-1}{\bm E}^{\rm T} \end{bmatrix}
\end{align}
where each diagonal block is a non-negative diagonal matrix. 
Considering the significant magnitude difference between the diagonal blocks ${\bm A}$ and ${\bm D}$ of ${\bm \varPhi}$, to neglect the perturbation requires, 
\begin{equation}
\begin{aligned}
   &{\rm Tr}({\bm G}{\bm F}^{-1}{\bm G}^{\rm T})\ll  {\rm Tr}({\bm A})\\
   &{\rm Tr}({\bm E}{\bm F}^{-1}{\bm E}^{\rm T})\ll  {\rm Tr}({\bm D}) 
\end{aligned}\label{eq:sufficient condition}
\end{equation}

According to \eqref{eq:g re}-\eqref{eq:FO moment squared mag}, we have 
\begin{equation}
   \small
\begin{aligned}
   {\rm Tr}({\bm G}{\bm F}^{-1}{\bm G}^{\rm T}) &=8\pi^2S\delta\sum_{n=1}^N\sum_{k=1}^K\rho_n|\alpha_{n,k}|^2\bar{f}_n^2\\
   {\rm Tr}({\bm E}{\bm F}^{-1}{\bm E}^{\rm T}) &=8\pi^2S\delta\sum_{n=1}^N\sum_{k=1}^K\rho_n|\alpha_{n,k}|^2(\bar{t}_n+\tau_{n,k})^2
\end{aligned}
\end{equation}
where $\bar{f}_n$ and ${\bar{t}_n}$ are defined in \eqref{eq:average frequency} and \eqref{eq:average time}, respectively.
In light of \eqref{eq:simplified SO-PD delay}-\eqref{eq:simplified SO-PD Doppler}, we have
\begin{equation}
   \small
\begin{aligned}
   {\rm Tr}({\bm A}) &= 8\pi^2S\delta\sum_{n=1}^N\sum_{k=1}^K\rho_n|\alpha_{n,k}|^2\bar{f_n^2},\\
   {\rm Tr}({\bm D}) &= 8\pi^2S\delta\sum_{n=1}^N\sum_{k=1}^K\rho_n|\alpha_{n,k}|^2\bar{t_n^2},
\end{aligned}
\end{equation}
where $\bar{f_n^2}$ and $\bar{t_n^2}$ are defined in \eqref{eq:square bandwidth} and \eqref{eq:square time}, respectively. Therefore, a sufficient condition for the tightness of the approximate CRLBs is $\bar{f}_n^2\ll \bar{f_n^2}$ and $(\bar{t}_n+\tau_{n,k})^2\ll \bar{t_n^2}$. Then, how to make it?

Firstly, according to Cauchy-Schwarz inequality \cite{Mitrinovic1993}, we have 
\begin{align}
   \bar{t}_n^2 &\leq \int t^2|s_n(t)|^2\ dt\int |s_n(t)|^2\ dt=\bar{t_n^2},\\
   {\bar{f}_n}^2 &\leq \int f^2|S_n(f)|^2\ df\int |S_n(f)|^2\ df=\bar{f_n^2}.
\end{align}
We now prove that $\bar{f_n^2}\propto  B_{\rm eff}$ and $\bar{t_n^2}\propto  T_{\rm eff}$.
Since $\int |S_n(f)|^2\ df=\int |s_n(t)|^2\ dt=1$, $|s_n(t)|^2$ and $|S_n(f)|^2$ can be interpreted as the probability density functions for consecutive random variables $t$ and $f$, respectively. Under this interpretation,  $\bar{t}_n$ and ${\bar{f}_n}$ represent the means, and the variances can be computed as follows 
\begin{align}
   {t}_n^2 &= \int (t-\bar{t}_n)^2|s_n(t)|^2\ dt = \bar{t}_n^2+\bar{t_n^2},\\
   {f}_n^2 &= \int (f-\bar{f}_n)^2|S_n(f)|^2\ df = \bar{f}_n^2+\bar{f_n^2}.
\end{align}
With the fixed time and frequency centers, increasing either the bandwidth or the pulse width results in a corresponding increase in the variances mentioned above, while the means remain nearly unchanged. In other words, with sufficiently large bandwidth and pulse time width, conditions such as $\bar{f}_n^2 \ll \bar{f_n^2}$ and $\bar{t}_n^2 \ll \bar{t_n^2}$ can be satisfied. 
Additionally, if the pulse time width is significantly greater than the delay $\tau_{n,k}$, we have $(\bar{t}_n + \tau_{n,k})^2 \ll \bar{t_n^2}$. Thus, a sufficient condition for ensuring the accuracy of the proposed approximate CRLBs is to utilize sensing waveforms with relatively large bandwidths and pulse time widths, satisfying the conditions $\bar{f}_n^2 \ll \bar{f_n^2}$ and $(\bar{t}_n + \tau_{n,k})^2 \ll \bar{t_n^2}$. \hfill$\blacksquare$
\section{CRLB for multiple-target sensing}\label{sec:CRLB multiple} 
In this section, we present detailed derivations of the CRLBs for multiple-target scenarios, extending the analysis provided in Section~\ref{sec:MT CRLB}. For the readability, the LLF in \eqref{eq:LF multiple} is rewritten here as
\begin{equation}
   \small
\begin{aligned}
   \log p({\bm r}|{\bm \phi})&= \frac{2}{\sigma_{z}^2}\sum_{n=1}^N\sum_{k=1}^Kf_s\mathfrak{R}\left\{ {\bm \alpha}_{n,k}^{\rm H}\int{r}_{n,k}(t){\bm y}_{n,k}^*(t)\, dt\right\}\\ 
   &-\frac{1}{\sigma_{z}^2}\sum_{n=1}^N\sum_{k=1}^Kf_s{\bm \alpha}_{n,k}^{\rm T}\int {\bm y}_{n,k}(t){\bm y}_{n,k}^{\rm H}(t)\, dt{\bm \alpha}_{n,k}^{*}.
\end{aligned}\label{eq:LLF multiple}
\end{equation}
To calculate ${\bm J}_{qq}$, we simplify \eqref{eq:LLF multiple} regarding any target $q$ as, 
\begin{align}
   \log p({\bm r}|{\bm \phi}_q)&= \frac{2f_s}{\sigma_{z}^2}\sum_{n=1}^N\sum_{k=1}^K\mathfrak{R}\left\{ {\alpha}_{n,k,q}\int{r}_{n,k}^{*}(t){y}_{n,k,q}(t)\, dt\right\}\notag\\ 
   &-\frac{f_s}{\sigma_{z}^2}\sum_{n=1}^N\sum_{k=1}^KP\rho_{n}|{\alpha}_{n,k,q}|^2,\label{eq:LLF multiple q}
\end{align}
Equation \eqref{eq:LLF multiple q} is nearly identical to \eqref{eq:LLF}, except that the received signal ${r}_{n,k}(t)$ now contains reflected signals from multiple targets, as indicated in \eqref{eq:received reflected sig}. Therefore, ${\bm J}_{qq}$ can be computed similarly to the single-target FIM in \eqref{eq:FIM_phi}.

The FIM block ${\bm J}_{ql}$ in \eqref{eq:FIM theta multiple} is computed by primarily using the second term in \eqref{eq:LLF multiple}. For simplification, we define the cross-correlation matrix among the $Q$ targets as follows:
\begin{align}
   {\bm Y}_{n,k} \triangleq  \int {\bm y}_{n,k}(t){\bm y}_{n,k}^{\rm H}(t)\, dt. \label{eq:cross-correlation}
\end{align}
Its $q$th diagonal element is ${Y}_{n,k,q,q}=P\rho_{n}$, and its non-diagonal element is given by,
\begin{align}
   {Y}_{n,k,q,l} &\triangleq  \int {y}_{n,k,q}(t){y}_{n,k,l}^{*}(t)\, dt \label{eq:auto-corre function}.
\end{align}
Since ${y}_{n,k,l}(t)$ is the corresponding time-frequency delayed version of ${y}_{n,k,q}(t)$, ${Y}_{n,k,q,l}$ can be interpreted as the time-frequency auto-correlation function of ${y}_{n,k,q}(t)$ with a time delay of $\tau_{n,k,q,l}$ and a frequency shift of $f_{n,k,q,l}$. Its squared modulus can be seen as the mutual ambiguity function between the corresponding pair of targets. We then extract the fundamental component of the second term in \eqref{eq:LLF multiple} for all $n,k,q,l$ and define it as
\begin{align}
   \vartheta_{n,k,q,l}\triangleq -\frac{2f_s}{\sigma_z^2}\mathfrak{R} \{\alpha_{n,k,q}{Y}_{n,k,q,l}\alpha_{n,k,l}^*\}.
\end{align}
Since ${Y}_{n,k,l,q}={Y}_{n,k,q,l}^*$, it follows that $\vartheta_{n,k,l,q}=\vartheta_{n,k,q,l}$. Then, the upper-triangular elements of ${\bm J}_{ql}$ for $q\neq l$ can be computed using \eqref{eq:simplified SO-PD delay}-\eqref{eq:F elements}, replacing the numerator $\log p({\bm r}|{\bm \phi})$ with $\vartheta_{n,k,q,l}$, and substituting the denominator with the corresponding time delay or Doppler frequency of the target pair $\{q,l\}$, e.g. $a_{n,k,q,l} = -\mathbb{E}_{{\bm r}}\left\{\frac{\partial^2 \vartheta_{n,k,q,l}}{\partial{\tau_{n,k,q}}\partial{\tau_{n,k,l}}}\right\}$.
The lower-triangular elements of ${\bm J}_{ql}$ can be similarly derived using the definition provided in \eqref{eq:FIM_phi q}.

\section{Proof of \textbf{Theorem 5}}\label{sec: proof of theorem 5}

We first rewrite ${\bm Q} = {\bm \aleph}{\bm \varPhi} {\bm \aleph}^{\rm T}$ as
\begin{align}
   {\bm Q}&=\begin{bmatrix}
      {\bm P}&{\bm V}\\ {\bm V}^{\rm T}&{\bm Y}
   \end{bmatrix} \label{eq:FIM theta simplified}
\end{align}
with its $2\times 2$ sub-matrices respectively given by
\begin{align}
   {\bm P}&={\bm \aleph}_{11}{\bm A}{\bm \aleph}_{11}^{\rm T}+{\bm \aleph}_{11}{\bm B}{\bm \aleph}_{12}^{\rm T}+{\bm \aleph}_{12}{\bm B}{\bm \aleph}_{11}^{\rm T}+{\bm \aleph}_{12}{\bm D}{\bm \aleph}_{12}^{\rm T}, \label{eq: mat P}
\end{align}
\begin{align}
{\bm V}&={\bm \aleph}_{11}{\bm B}{\bm \aleph}_{22}^{\rm T}+{\bm \aleph}_{12}{\bm D}{\bm \aleph}_{22}^{\rm T},\label{eq: mat V}
\end{align}
and
\begin{align}
{\bm Y}&={\bm \aleph}_{22}{\bm D}{\bm \aleph}_{22}^{\rm T}.\label{eq: mat Y}
\end{align}
Taking the inversion of ${\bm Q}$ yields,
\begin{align}
   {\bm Q}^{-1}&=\begin{bmatrix}
      C_{\rm L}&{\bm \varOmega}\\ {\bm \varOmega}&C_{\rm V}
   \end{bmatrix}.
\end{align}
According to the blockwise matrix inversion property in \cite{MatrixCookbook2012}, we have the CRLB matrix for location estimation,
\begin{align}
   C_{\rm L}& = ({\bm P}-{\bm V}{\bm Y}^{-1}{\bm V}^{\rm T})^{-1}, \label{eq: CRLB location}
\end{align}
and the CRLB matrix for velocity estimation,
\begin{align}
   C_{\rm V} &= ({\bm Y}-{\bm V}^{\rm T}{\bm P}^{-1}{\bm V})^{-1}.\label{eq: CRLB velocity}
\end{align}

Define the equivalent location FIM from \eqref{eq: CRLB location} as
\begin{align}
   J_{\rm L}& = {\bm P}-{\bm V}{\bm Y}^{-1}{\bm V}^{\rm T}. \label{eq:location FIM}
\end{align}
With a band-limited, Gaussian transmission waveform for sensing, we have 
\begin{align}
{\bm \aleph}_{11}{\bm B}{\bm \aleph}_{12}^{\rm T} &= {\bm \aleph}_{11}{\bm B}{\bm \aleph}_{22}^{\rm T}{\bm Y}^{-1}({\bm \aleph}_{12}{\bm D}{\bm \aleph}_{22}^{\rm T})^{\rm T},\label{eq: equality 1}\\
{\bm \aleph}_{11}{\bm A}{\bm \aleph}_{11}^{\rm T} &= {\bm \aleph}_{11}{\bm B}{\bm \aleph}_{22}^{\rm T}{\bm Y}^{-1}({\bm \aleph}_{11}{\bm B}{\bm \aleph}_{22}^{\rm T})^{\rm T}. \label{eq: equality 2}
\end{align} 
Therefore, \eqref{eq:location FIM} can be further simplified as,
\begin{align}
   J_{\rm L}& = {\bm \aleph}_{12}{\bm D}{\bm \aleph}_{12}^{\rm T}-{\bm \aleph}_{12}{\bm D}{\bm \aleph}_{22}^{\rm T}{\bm Y}^{-1}{\bm \aleph}_{22}{\bm D}{\bm \aleph}_{12}^{\rm T}. \label{eq:simplify location FIM}
\end{align}
We observe that the FIM $J_{\rm L}$ depends on the matrix ${\bm D}$ as defined in \eqref{eq:simplified SO-PD Doppler}. Consequently, the location CRLBs are inversely proportional to the SETW, but are largely insensitive to the SEBW.

We now briefly prove \eqref{eq: equality 1} and \eqref{eq: equality 2}. 
Firstly, according to Cauchy-Schwarz inequality \cite{Mitrinovic1993}, we have, 
\begin{align}
\bar{f_n^2}\bar{t_n^2}&\geq \frac{1}{4\pi^2}|\sigma_{tf}|^2\notag\\
&=\frac{1}{4\pi^2}(\mathfrak{R}^2(\sigma_{tf})+\mathfrak{T}^2(\sigma_{tf}))\notag\\
&=\frac{1}{4\pi^2}(1/4+\mathfrak{T}^2(\sigma_{tf}))
\end{align}
where the equality holds for a band-limited, Gaussian transmission waveform $s(t)$, e.g., single Gaussian pulse shaped OCDM in \eqref{eq:OCDM}. Furthermore, considering a large SEBW $\bar{f_n^2}$ and SETW $\bar{t_n^2}$ to enable $\mathfrak{T}^2(\sigma_{tf})\gg 1/4$, we have 
\begin{align}
   \bar{f_n^2}\bar{t_n^2}\approx\frac{1}{4\pi^2}\mathfrak{T}^2(\sigma_{tf}),
\end{align}
or equivalently, ${\bm B}^2={\bm A}{\bm D}$. According to their element-wise relations, we further have ${\bm B}=c_a {\bm A}$ and ${\bm B}=c_d {\bm D}$ with $c_ac_d=1$.

Secondly, the relevant geometric spread parameters are presented for any $1\leq n\leq N$ and $1\leq k\leq K$, i.e., 
\begin{align}
   \begin{cases}
   \beta_{n,k} = \dfrac{1}{c}\left(\dfrac{x-x_n}{\|{\bm l}_n-{\bm l}\|_2}+\dfrac{x-x_k}{\|{\bm l}_k-{\bm l}\|_2}\right),\\
   \zeta_{n,k} = \dfrac{1}{c}\left(\dfrac{y-y_n}{\|{\bm l}_n-{\bm l}\|_2}+\dfrac{y-y_k}{\|{\bm l}_k-{\bm l}\|_2}\right),\\
   \xi_{n,k}= \dfrac{f_c}{c}\left(\dfrac{x_n-x}{\|{\bm l}_n-{\bm l}\|_2}+\dfrac{x_k-x}{\|{\bm l}_k-{\bm l}\|_2}\right),\\
   \varrho_{n,k}= \dfrac{f_c}{c}\left(\dfrac{y_n-y}{\|{\bm l}_n-{\bm l}\|_2}+\dfrac{y_k-y}{\|{\bm l}_k-{\bm l}\|_2}\right).
   \end{cases}\label{eq:GeoSpread para}
\end{align}
It is evident that $\xi_{n,k}=-f_c\beta_{n,k}$ and $\varrho_{n,k}=-f_c\zeta_{n,k}$ for each pair of $\{n,k\}$, and thus ${\bm \aleph}_{22}=-f_c{\bm \aleph}_{11}$.

Then, we can convert \eqref{eq: equality 1} to
\begin{align}
   &{\bm \aleph}_{11}{\bm B}{\bm \aleph}_{12}^{\rm T} = {\bm \aleph}_{11}{\bm B}{\bm \aleph}_{22}^{\rm T}{\bm Y}^{-1}({\bm \aleph}_{12}{\bm D}{\bm \aleph}_{22}^{\rm T})^{\rm T}\notag\\
   &\Leftrightarrow {\bm \aleph}_{11}{\bm B} ({\bm I}-{\bm \aleph}_{22}^{\rm T}{\bm Y}^{-1}{\bm \aleph}_{22}{\bm D}){\bm \aleph}_{12}^{\rm T}={\bm O}\notag\\
   &\Leftrightarrow \frac{-c_d}{f_c}{\bm \aleph}_{22}{\bm D}({\bm I}-{\bm \aleph}_{22}^{\rm T}{\bm Y}^{-1}{\bm \aleph}_{22}{\bm D}){\bm \aleph}_{12}^{\rm T}={\bm O}.\label{eq:prove equality 1}
\end{align}
Since ${\bm I}-{\bm \aleph}_{22}^{\rm T}{\bm Y}^{-1}{\bm \aleph}_{22}{\bm D}$ is a projection matrix onto ${\bm \aleph}_{22}{\bm D}$, 
\eqref{eq:prove equality 1} is proved.
On the other hand, \eqref{eq: equality 2} can be converted to
\begin{align}
   &{\bm \aleph}_{11}{\bm A}{\bm \aleph}_{11}^{\rm T} = {\bm \aleph}_{11}{\bm B}{\bm \aleph}_{22}^{\rm T}{\bm Y}^{-1}({\bm \aleph}_{11}{\bm B}{\bm \aleph}_{22}^{\rm T})^{\rm T}\notag\\
   &\Leftrightarrow \frac{-1}{f_c}{\bm \aleph}_{22}{\bm A}({\bm I}-{\bm \aleph}_{22}^{\rm T}({\bm \aleph}_{22}{\bm A}{\bm \aleph}_{22}^{\rm T})^{-1}{\bm \aleph}_{22}{\bm A}){\bm \aleph}_{11}^{\rm T}={\bm O}. \label{eq:prove equality 2}
\end{align}
Similarly, ${\bm I}-{\bm \aleph}_{22}^{\rm T}({\bm \aleph}_{22}{\bm A}{\bm \aleph}_{22}^{\rm T})^{-1}{\bm \aleph}_{22}{\bm A}$ is a projection matrix onto ${\bm \aleph}_{22}{\bm A}$, and thus \eqref{eq:prove equality 2} holds. Therefore, \eqref{eq: equality 1} and \eqref{eq: equality 2} are proved. \hfill$\blacksquare$

\section{Proof of \textbf{Theorem 6}}\label{sec:CRLB trend}
As analyzed in Section~\ref{sec:MT CRLB} and Appendix~\ref{sec:CRLB multiple}, each diagonal block of the FIM in \eqref{eq:FIM theta multiple} incorporates contributions from the reflected signals of all targets, resulting in additive coupling between the target of interest and the remaining targets. Additionally, the off-diagonal FIM blocks in \eqref{eq:FIM theta multiple} represent direct structural coupling between each pair of targets, arising from their mutual cross-correlation functions.

To proceed, let $[{\bm C}]_q$ denote the CRLB for target $q$ in a multi-target sensing scenario. A straightforward method to simplify the original FIM is to approximate it by a block-diagonal matrix, effectively eliminating the structural coupling between different targets. This can be achieved by neglecting each off-diagonal block ${\bm J}_{ql}$ in \eqref{eq:FIM theta multiple}, yielding an approximate but decoupled CRLB estimate for each target, i.e.,
\begin{align}
   \tilde{\bm C}_q = ({\bm \varLambda}_q{\bm J}_{qq}{\bm \varLambda}_q)^{-1}. \label{eq:CRLB target q}
\end{align}   

The additive coupling in ${\bm J}_{qq}$ comes from the reflected signals of multiple targets, as described in \eqref{eq:received reflected sig}. After cancelling the additive coupling, $[{\bm C}]_q$ finally reduce to the counterpart in corresponding single-target sensing scenario, i.e,
\begin{align}
   {\bm C}_q = ({\bm \varLambda}_q{\bm J}_{q}{\bm \varLambda}_q)^{-1}, \label{eq:approximate CRLB target}
\end{align}   
where ${\bm J}_{q}$ denotes the FIM for target $q$ in \eqref{eq:FIM_phi}.

To enable the approximate CRLB in \eqref{eq:CRLB target q}, each non-diagonal FIM block ${\bm J}_{ql}$ should satisfy $\|{\bm J}_{ql}\|_2\ll  \|{\bm J}_{qq}\|_2$ and $\|{\bm J}_{ql}\|_2\ll  \|{\bm J}_{ll}\|_2$. \emph{To this end, ${Y}_{n,k,q,l}$, and its first- and second-order partial derivatives regarding either delay or Doppler frequency, should be relatively small.} The Doppler frequency resolution and the time delay resolution are denoted as $f_{\rm r}$ and $\tau_{\rm r}$, respectively. We generally have $\tau_{\rm r}\leq T_{\rm eff}$, $f_{\rm r}\leq B_{\rm eff}$ and $f_{\rm r}\propto 1/\tau_{\rm r}$. Let us rewrite \eqref{eq:auto-corre function} as,
\begin{align}
   {Y}_{n,k,q,l} =\ &P\rho_n\int s_{n}(t-\tau_{n,k,q})s_n^{*}(t-\tau_{n,k,l})e^{j\omega_{n,k,q,l}t}\, dt,\label{eq:auto-corre function time}\\
      =\ &\frac{P\rho_n}{2\pi} e^{j\omega_{n,k,q,l}\tau_{n,k,q}}\notag\\
      &\cdot \int S_{n}(\omega-\omega_{n,k,q,l})S_n^{*}(\omega)e^{-j\omega \tau_{n,k,q,l}}\, d \omega, \label{eq:auto-corre function frequency}
\end{align}
with the relative Doppler angle frequency $\omega_{n,k,q,l} \triangleq 2\pi f_{n,k,q,l}$.
When $|\tau_{n,k,q,l}|\geq \tau_{\rm r}$ or $|f_{n,k,q,l}|\geq f_{\rm r}$ over widely separated targets in delay-Doppler domain, the corresponding time-frequency auto-correlation function modulus, $|{Y}_{n,k,q,l}|$, will approach zero. This point is also mentioned in \cite{van2001detection}. 

Define $\dot{s}_n(t)$ and $\ddot{s}_n(t)$ as the first- and second-order derivatives of ${s}_n(t)$ regarding $t$, respectively, while the corresponding frequency-domain derivatives are denoted by $\dot{S}_n(\omega)$ and $\ddot{S}_n(\omega)$. By respectively taking the first-order derivative of ${Y}_{n,k,q,l}$ regarding $\tau_{n,k,q}$ and $f_{n,k,q}$, we have 
\footnotesize
\begin{align}
   \frac{\partial {Y}_{n,k,q,l}}{\partial{\tau_{n,k,q}}} =\ & -P\rho_n\int \dot{s}_{n}(t-\tau_{n,k,q})s_n^{*}(t-\tau_{n,k,l})e^{j\omega_{n,k,q,l}t}\, dt,\label{eq:FODT auto-corre function time}\\
      =\ & -j\dfrac{P\rho_n}{2\pi} e^{j\omega_{n,k,q,l}\tau_{n,k,q}}\notag\\
      &\cdot \int (\omega-\omega_{n,k,q,l})S_{n}(\omega-\omega_{n,k,q,l})S_n^{*}(\omega)e^{-j\omega\tau_{n,k,q,l}}\, d \omega, \label{eq:FODT auto-corre function frequency}
\end{align}
\normalsize
and 
\footnotesize
\begin{align}
   \frac{\partial {Y}_{n,k,q,l}}{\partial{f_{n,k,q}}} =\ & j2\pi P\rho_n\int t{s}_{n}(t-\tau_{n,k,q})s_n^{*}(t-\tau_{n,k,l})e^{j\omega_{n,k,q,l}t}\, dt\label{eq:FODF auto-corre function time}\\
      =\ & jP\rho_n e^{j\omega_{n,k,q,l}\tau_{n,k,l}} \int \left(j\dot{S}_n(\omega)+\tau_{n,k,q}S_n(\omega)\right)\notag\\
      &\cdot S_{n}^*(\omega+\omega_{n,k,q,l})e^{-j\omega\tau_{n,k,q,l}}\, d \omega. \label{eq:FODF auto-corre function frequency}
\end{align}
\normalsize
\begin{figure*}[!t]
   \footnotesize
\begin{align}
   \frac{\partial^2 {Y}_{n,k,q,l}}{\partial{\tau_{n,k,q}}\partial{\tau_{n,k,l}}} &= P\rho_n\int \dot{s}_{n}(t-\tau_{n,k,q})\dot{s}_n^{*}(t-\tau_{n,k,l})e^{j\omega_{n,k,q,l}t}\, dt,\label{eq:SODT auto-corre function time}\\
      &= \dfrac{P\rho_n}{2\pi} e^{j\omega_{n,k,q,l}\tau_{n,k,q}}\int \omega(\omega-\omega_{n,k,q,l})S_{n}(\omega-\omega_{n,k,q,l})S_n^{*}(\omega)e^{-j\omega\tau_{n,k,q,l}}\, d \omega, \label{eq:SODT auto-corre function frequency}\\
   \frac{\partial^2 {Y}_{n,k,q,l}}{\partial{f_{n,k,q}}\partial{f_{n,k,l}}} &= 4\pi^2 P\rho_n\int t^2{s}_{n}(t-\tau_{n,k,q})s_n^{*}(t-\tau_{n,k,l})e^{j\omega_{n,k,q,l}t}\, dt,\label{eq:SODF auto-corre function time}\\
      &= 2\pi P\rho_n e^{j\omega_{n,k,q,l}\tau_{n,k,l}} \int \left(\ddot{S}_n(\omega)-2j\tau_{n,k,q}\dot{S}_n(\omega)-\tau^2_{n,k,q}S_n(\omega)\right)S_{n}^*(\omega+\omega_{n,k,q,l})e^{-j\omega\tau_{n,k,q,l}}\, d \omega, \label{eq:SODF auto-corre function frequency}\\
   \frac{\partial^2 {Y}_{n,k,q,l}}{\partial{\tau_{n,k,q}}\partial{f_{n,k,l}}} &= j2\pi P\rho_n\int t\dot{s}_{n}(t-\tau_{n,k,q})s_n^{*}(t-\tau_{n,k,l})e^{j\omega_{n,k,q,l}t}\, dt,\label{eq:SOTF auto-corre function time}\\
   &= P\rho_n e^{j\omega_{n,k,q,l}\tau_{n,k,q}} \int (\omega-\omega_{n,k,q,l})\left(j\dot{S}_n^*(\omega)-\tau_{n,k,l}S_n^*(\omega)\right)S_{n}(\omega+\omega_{n,k,q,l})e^{-j\omega\tau_{n,k,q,l}}\, d \omega. \label{eq:SOTF auto-corre function frequency}
\end{align}\end{figure*}
Similarly, the second-order partial derivatives of ${Y}_{n,k,q,l}$ are, respectively, given by \eqref{eq:SODT auto-corre function time}-\eqref{eq:SOTF auto-corre function frequency}.
We can observe from \eqref{eq:FODT auto-corre function time}-\eqref{eq:SOTF auto-corre function frequency} that, with $|\tau_{n,k,q,l}|\geq \tau_{\rm r}$ or $|f_{n,k,q,l}|\geq f_{\rm r}$, both the first- and second-order partial derivatives approach zero.
\emph{Therefore, a sufficient condition for achieving the approximate CRLB in \eqref{eq:CRLB target q} is that all targets are widely separated in delay and Doppler domain with $|\tau_{n,k,q,l}|\geq \tau_{\rm r}$ or $|f_{n,k,q,l}|\geq f_{\rm r}$.}

On the other hand, the additive coupling component in \eqref{eq:LLF multiple q} due to any other target $l$ is given by,
\begin{equation}
   \small
   \begin{aligned}
   i_{n,k,q,l}\triangleq\ &\int{y}_{n,k,l}^{*}(t){y}_{n,k,q}(t)\, dt\\
   =\ &P\rho_n\int s_{n}(t-\tau_{n,k,q})s_n^{*}(t-\tau_{n,k,l})e^{j\omega_{n,k,q,l}t}\, dt,
   \end{aligned}\label{eq:interf component}
\end{equation}
where the noise component has been neglected by taking the mathematical expectation for ${r}_{n,k}(t)$ when calculating the CRLB. We can observe that $i_{n,k,q,l}$ in \eqref{eq:interf component} is the same as ${Y}_{n,k,q,l}$ in \eqref{eq:auto-corre function time}, and the same holds for their first-order partial derivatives. Its second-order partial derivatives regarding the delay and Doppler frequency are respectively given by,
\begin{align}
   &\frac{\partial^2 i_{n,k,q,l}}{\partial{\tau_{n,k,q}^2}} = P\rho_n\int \ddot{s}_{n}(t-\tau_{n,k,q}){s}_n^{*}(t-\tau_{n,k,l})e^{j\omega_{n,k,q,l}t}\, dt\label{eq:SODT auto-corre function time q}\\
      &~~~~~~= - \dfrac{P\rho_n}{2\pi} e^{j\omega_{n,k,q,l}\tau_{n,k,q}}\label{eq:SODT auto-corre function frequency q}\\
      &~~~~~~\cdot \int (\omega-\omega_{n,k,q,l})^2S_{n}(\omega-\omega_{n,k,q,l})S_n^{*}(\omega)e^{-j\omega\tau_{n,k,q,l}}\, d \omega\notag 
\end{align}
\begin{align}
   \frac{\partial^2 i_{n,k,q,l}}{\partial{f_{n,k,q}^2}} &= -\frac{\partial^2 Y_{n,k,q,l}}{\partial{f_{n,k,q}}\partial{f_{n,k,l}}}
\end{align}
and 
\begin{align}
   \frac{\partial^2 i_{n,k,q,l}}{\partial{\tau_{n,k,q}}\partial{f_{n,k,q}}} &= -\frac{\partial^2 Y_{n,k,q,l}}{\partial{\tau_{n,k,q}}\partial{f_{n,k,l}}}.\label{eq:SOTF auto-corre function time q}
\end{align}

According to \eqref{eq:auto-corre function time}-\eqref{eq:SOTF auto-corre function frequency} and \eqref{eq:SODT auto-corre function time q}-\eqref{eq:SOTF auto-corre function time q}, we know that with $|\tau_{n,k,q,l}|\geq \tau_{\rm r}$ or $|f_{n,k,q,l}|\geq f_{\rm r}$, both the additive coupling component \eqref{eq:interf component} and its first- and second-order partial derivatives regarding the delay or Doppler frequency of target $q$ approach zero. Consequently, a sufficient condition for the multiple-target CRLB achieving the single-target CRLB in \eqref{eq:approximate CRLB target} is the same as that for achieving \eqref{eq:CRLB target q}, and we have $\tilde{\bm C}_q={\bm C}_q$ under this condition. \hfill$\blacksquare$

\small
\bibliographystyle{IEEEtran}
\bibliography{mybibfile_ISAC.bib}

\begin{thebibliography}{10}
\providecommand{\url}[1]{#1}
\csname url@samestyle\endcsname
\providecommand{\newblock}{\relax}
\providecommand{\bibinfo}[2]{#2}
\providecommand{\BIBentrySTDinterwordspacing}{\spaceskip=0pt\relax}
\providecommand{\BIBentryALTinterwordstretchfactor}{4}
\providecommand{\BIBentryALTinterwordspacing}{\spaceskip=\fontdimen2\font plus
\BIBentryALTinterwordstretchfactor\fontdimen3\font minus
  \fontdimen4\font\relax}
\providecommand{\BIBforeignlanguage}[2]{{%
\expandafter\ifx\csname l@#1\endcsname\relax
\typeout{** WARNING: IEEEtran.bst: No hyphenation pattern has been}%
\typeout{** loaded for the language `#1'. Using the pattern for}%
\typeout{** the default language instead.}%
\else
\language=\csname l@#1\endcsname
\fi
#2}}
\providecommand{\BIBdecl}{\relax}
\BIBdecl

\bibitem{Pang2024}
X.~Pang, S.~Guo, J.~Tang, N.~Zhao, and N.~Al-Dhahir, ``Dynamic {ISAC}
  beamforming design for {UAV}-enabled vehicular networks,'' \emph{IEEE Trans.
  Wireless Commun.}, vol.~23, no.~11, pp. 16\,852--16\,864, 2024.

\bibitem{Nuria2024}
N.~González-Prelcic, M.~Furkan~Keskin, O.~Kaltiokallio, M.~Valkama,
  D.~Dardari, X.~Shen, Y.~Shen, M.~Bayraktar, and H.~Wymeersch, ``The
  integrated sensing and communication revolution for {6G}: Vision, techniques,
  and applications,'' \emph{Proc. IEEE}, vol. 112, no.~7, pp. 676--723, 2024.

\bibitem{Liu2022survey}
A.~Liu, Z.~Huang, M.~Li, Y.~Wan, W.~Li, T.~X. Han, C.~Liu, R.~Du, D.~K.~P. Tan,
  J.~Lu, Y.~Shen, F.~Colone, and K.~Chetty, ``A survey on fundamental limits of
  integrated sensing and communication,'' \emph{IEEE Commun. Surveys {\&}
  Tutorials}, vol.~24, no.~2, pp. 994--1034, 2022.

\bibitem{Xiong2023}
Y.~Xiong, F.~Liu, Y.~Cui, W.~Yuan, T.~X. Han, and G.~Caire, ``On the
  fundamental tradeoff of integrated sensing and communications under
  {Gaussian} channels,'' \emph{IEEE Trans. Inf. Theory}, vol.~69, no.~9, pp.
  5723--5751, 2023.

\bibitem{Liu2022}
F.~Liu, Y.~Cui, C.~Masouros, J.~Xu, T.~X. Han, Y.~C. Eldar, and S.~Buzzi,
  ``Integrated sensing and communications: Toward dual-functional wireless
  networks for {6G} and beyond,'' \emph{IEEE J. Sel. Areas Commun.}, vol.~40,
  no.~6, pp. 1728--1767, 2022.

\bibitem{Haimovich2008}
A.~M. Haimovich, R.~S. Blum, and L.~J. Cimini, ``{MIMO} radar with widely
  separated antennas,'' \emph{IEEE Signal Process. Mag.}, vol.~25, no.~1, pp.
  116--129, 2008.

\bibitem{Ganesan2021}
U.~K. Ganesan, E.~Björnson, and E.~G. Larsson, ``Clustering-based activity
  detection algorithms for grant-free random access in cell-free massive
  {MIMO},'' \emph{IEEE Trans. Commun.}, vol.~69, no.~11, pp. 7520--7530, 2021.

\bibitem{Guo2022}
M.~Guo and M.~C. Gursoy, ``Joint activity detection and channel estimation in
  cell-free massive {MIMO} networks with massive connectivity,'' \emph{IEEE
  Trans. Commun.}, vol.~70, no.~1, pp. 317--331, 2022.

\bibitem{Ammar2022Cell_free}
H.~A. Ammar, R.~Adve, S.~Shahbazpanahi, G.~Boudreau, and K.~V. Srinivas,
  ``Downlink resource allocation in multiuser cell-free {MIMO} networks with
  user-centric clustering,'' \emph{IEEE Trans. Wireless Commun.}, vol.~21,
  no.~3, pp. 1482--1497, 2022.

\bibitem{Behdad2022}
Z.~Behdad, O.~T. Demir, K.~W. Sung, E.~Björnson, and C.~Cavdar, ``Power
  allocation for joint communication and sensing in cell-free massive {MIMO},''
  in \emph{GLOBECOM 2022-2022 IEEE Global Commun. Conf.}, 2022, pp. 4081--4086.

\bibitem{Behdad2024}
------, ``Multi-static target detection and power allocation for integrated
  sensing and communication in cell-free massive {MIMO},'' \emph{IEEE Trans.
  Wireless Commun.}, pp. 1--1, 2024.

\bibitem{Ahmed2019}
A.~Ahmed, Y.~D. Zhang, and B.~Himed, ``Distributed dual-function
  radar-communication {MIMO} system with optimized resource allocation,'' in
  \emph{2019 IEEE Radar Conf. (RadarConf)}, 2019, pp. 1--5.

\bibitem{Huang2023}
Y.~Huang, Y.~Fang, X.~Li, and J.~Xu, ``Coordinated power control for network
  integrated sensing and communication,'' \emph{IEEE Trans. Veh. Technol.},
  vol.~71, no.~12, pp. 13\,361--13\,365, 2022.

\bibitem{Ahmed2024}
A.~Ahmed and Y.~D. Zhang, ``Optimized resource allocation for distributed joint
  radar-communication system,'' \emph{IEEE Trans. Veh. Technol.}, vol.~73,
  no.~3, pp. 3872--3885, 2024.

\bibitem{Weihao2024}
W.~Mao, Y.~Lu, C.-Y. Chi, B.~Ai, Z.~Zhong, and Z.~Ding, ``Communication-sensing
  region for cell-free massive {MIMO ISAC} systems,'' \emph{IEEE Trans.
  Wireless Commun.}, vol.~23, no.~9, pp. 12\,396--12\,411, 2024.

\bibitem{Tajer2010}
A.~Tajer, G.~H. Jajamovich, X.~Wang, and G.~V. Moustakides, ``Optimal joint
  target detection and parameter estimation by {MIMO} radar,'' \emph{IEEE J.
  Sel. Topics Signal Process.}, vol.~4, no.~1, pp. 127--145, 2010.

\bibitem{Fang2023}
Y.~Fang, S.~Zhu, B.~Liao, X.~Li, and G.~Liao, ``Target localization with
  bistatic {MIMO} and {FDA-MIMO} dual-mode radar,'' \emph{IEEE Trans. Aerosp.
  Electron. Syst.}, vol.~60, no.~1, pp. 952--964, 2024.

\bibitem{An2023}
J.~An, H.~Li, D.~W.~K. Ng, and C.~Yuen, ``Fundamental detection probability vs.
  achievable rate tradeoff in integrated sensing and communication systems,''
  \emph{IEEE Trans. Wireless Commun.}, vol.~22, no.~12, pp. 9835--9853, 2023.

\bibitem{Dwivedi2018}
S.~Dwivedi, P.~Aggarwal, and A.~K. Jagannatham, ``Fast block {LMS} and
  {RLS}-based parameter estimation and two-dimensional imaging in monostatic
  {MIMO} radar systems with multiple mobile targets,'' \emph{IEEE Trans. Signal
  Process.}, vol.~66, no.~7, pp. 1775--1790, 2018.

\bibitem{Qi2022}
Q.~Qi, X.~Chen, A.~Khalili, C.~Zhong, Z.~Zhang, and D.~W.~K. Ng, ``Integrating
  sensing, computing, and communication in {6G} wireless networks: Design and
  optimization,'' \emph{IEEE Trans. Commun.}, vol.~70, no.~9, pp. 6212--6227,
  2022.

\bibitem{Dianat2013}
M.~Dianat, M.~R. Taban, J.~Dianat, and V.~Sedighi, ``Target localization using
  least squares estimation for {MIMO} radars with widely separated antennas,''
  \emph{IEEE Trans. Aerosp. Electron. Syst.}, vol.~49, no.~4, pp. 2730--2741,
  2013.

\bibitem{Park2015}
C.-H. Park and J.-H. Chang, ``Closed-form localization for distributed {MIMO}
  radar systems using time delay measurements,'' \emph{IEEE Trans. Wireless
  Commun.}, vol.~15, no.~2, pp. 1480--1490, 2016.

\bibitem{Chen2023}
Z.~Chen, J.~Tang, L.~Huang, Z.-Q. He, K.-K. Wong, and J.~Wang, ``Robust target
  positioning for reconfigurable intelligent surface assisted {MIMO} radar
  systems,'' \emph{IEEE Trans. Veh. Technol.}, vol.~72, no.~11, pp.
  15\,098--15\,102, 2023.

\bibitem{He2010Velocity}
Q.~He, R.~S. Blum, H.~Godrich, and A.~M. Haimovich, ``Target velocity
  estimation and antenna placement for {MIMO} radar with widely separated
  antennas,'' \emph{IEEE J. Sel. Topics Signal Process.}, vol.~4, no.~1, pp.
  79--100, 2010.

\bibitem{Wang2023}
M.~Wang, X.~Li, L.~Gao, Z.~Sun, G.~Cui, and T.~S. Yeo, ``Signal accumulation
  method for high-speed maneuvering target detection using airborne coherent
  {MIMO} radar,'' \emph{IEEE Trans. Signal Process.}, vol.~71, pp. 2336--2351,
  2023.

\bibitem{Boyer2011}
R.~Boyer, ``Performance bounds and angular resolution limit for the moving
  colocated {MIMO} radar,'' \emph{IEEE Trans. Signal Process.}, vol.~59, no.~4,
  pp. 1539--1552, 2011.

\bibitem{Liao2018}
B.~Liao, ``Fast angle estimation for {MIMO} radar with nonorthogonal
  waveforms,'' \emph{IEEE Trans. Aerosp. Electron. Syst.}, vol.~54, no.~4, pp.
  2091--2096, 2018.

\bibitem{Ouyang2023MI}
C.~Ouyang, Y.~Liu, H.~Yang, and N.~Al-Dhahir, ``Integrated sensing and
  communications: A mutual information-based framework,'' \emph{IEEE Commun.
  Mag.}, vol.~61, no.~5, pp. 26--32, 2023.

\bibitem{Mohammad2023}
M.~Al-Jarrah, E.~Alsusa, and C.~Masouros, ``A unified performance framework for
  integrated sensing-communications based on {KL}-divergence,'' \emph{IEEE
  Trans. Wireless Commun.}, vol.~22, no.~12, pp. 9390--9411, 2023.

\bibitem{Li2007}
J.~Li and P.~Stoica, ``{MIMO} radar with colocated antennas,'' \emph{IEEE
  Signal Process. Mag.}, vol.~24, no.~5, pp. 106--114, 2007.

\bibitem{Fishler2006}
E.~Fishler, A.~Haimovich, R.~Blum, L.~Cimini, D.~Chizhik, and R.~Valenzuela,
  ``Spatial diversity in radars—models and detection performance,''
  \emph{IEEE Trans. Signal Process.}, vol.~54, no.~3, pp. 823--838, 2006.

\bibitem{Wang2014TGRS}
W.-Q. Wang, ``{MIMO SAR OFDM} chirp waveform diversity design with random
  matrix modulation,'' \emph{IEEE Trans. Geoscience and Remote Sensing},
  vol.~53, no.~3, pp. 1615--1625, 2015.

\bibitem{Aldayel2016}
O.~Aldayel, V.~Monga, and M.~Rangaswamy, ``Successive {QCQP} refinement for
  {MIMO} radar waveform design under practical constraints,'' \emph{IEEE Trans.
  Signal Process.}, vol.~64, no.~14, pp. 3760--3774, 2016.

\bibitem{Xue2024}
B.~Xue, G.~Zhang, F.~Gini, M.~S. Greco, and H.~Leung, ``An optimized
  interleaved {OFDM} chirp orthogonal waveform design for dechirped miniature
  {MMW MIMO} radar,'' in \emph{2024 IEEE Int. Conf. Acoust., Speech and Signal
  Process. (ICASSP)}, 2024, pp. 8701--8705.

\bibitem{He2010}
Q.~He, R.~S. Blum, and A.~M. Haimovich, ``Noncoherent {MIMO} radar for location
  and velocity estimation: More antennas means better performance,'' \emph{IEEE
  Trans. Signal Process.}, vol.~58, no.~7, pp. 3661--3680, 2010.

\bibitem{Chuanming2010}
C.~Wei, Q.~He, and R.~S. Blum, ``Cramer-rao bound for joint location and
  velocity estimation in multi-target non-coherent {MIMO} radars,'' in
  \emph{2010 44th Ann. Conf. Inform. Sci. and Syst. (CISS)}, 2010, pp. 1--6.

\bibitem{he2012noncoherent}
Q.~He and R.~S. Blum, ``Noncoherent versus coherent {MIMO} radar: Performance
  and simplicity analysis,'' \emph{Signal Process.}, vol.~92, no.~10, pp.
  2454--2463, 2012.

\bibitem{He2016}
Q.~He, J.~Hu, R.~S. Blum, and Y.~Wu, ``Generalized cramér-rao bound for joint
  estimation of target position and velocity for active and passive radar
  networks,'' \emph{IEEE Trans. Signal Process.}, vol.~64, no.~8, pp.
  2078--2089, 2016.

\bibitem{Godrich2010}
H.~Godrich, A.~M. Haimovich, and R.~S. Blum, ``Target localization accuracy
  gain in {MIMO} radar-based systems,'' \emph{IEEE Trans. Inf. Theory},
  vol.~56, no.~6, pp. 2783--2803, 2010.

\bibitem{Ai2015}
Y.~Ai, W.~Yi, R.~S. Blum, and L.~Kong, ``Cramer-rao lower bound for multitarget
  localization with noncoherent statistical {MIMO} radar,'' in \emph{2015 IEEE
  Radar Conf. (RadarCon)}, 2015, pp. 1497--1502.

\bibitem{Godrich2012}
H.~Godrich, A.~P. Petropulu, and H.~V. Poor, ``Sensor selection in distributed
  multiple-radar architectures for localization: A knapsack problem
  formulation,'' \emph{IEEE Trans. Signal Process.}, vol.~60, no.~1, pp.
  247--260, 2012.

\bibitem{Godrich2011}
------, ``Power allocation strategies for target localization in distributed
  multiple-radar architectures,'' \emph{IEEE Trans. Signal Process.}, vol.~59,
  no.~7, pp. 3226--3240, 2011.

\bibitem{feng2016fast}
H.-Z. Feng, H.-W. Liu, J.-K. Yan, F.-Z. Dai, and M.~Fang, ``A fast efficient
  power allocation algorithm for target localization in cognitive distributed
  multiple radar systems,'' \emph{Signal Process.}, vol. 127, pp. 100--116,
  2016.

\bibitem{van2001detection}
H.~L. Van~Trees, \emph{Detection, estimation, and modulation theory}.\hskip 1em
  plus 0.5em minus 0.4em\relax John Wiley \& Sons, 2001.

\bibitem{Melvin2013ModernRadar}
W.~L. Melvin and J.~Scheer, \emph{Principles of modern radar. Vol. II, Advanced
  techniques}.\hskip 1em plus 0.5em minus 0.4em\relax Raleigh: SciTech Pub.,
  2013.

\bibitem{He2010Homogeneous}
Q.~He, N.~H. Lehmann, R.~S. Blum, and A.~M. Haimovich, ``{MIMO} radar moving
  target detection in homogeneous clutter,'' \emph{IEEE Trans. Aerosp.
  Electron. Syst.}, vol.~46, no.~3, pp. 1290--1301, 2010.

\bibitem{Wang2011Nonhomogeneous}
P.~Wang, H.~Li, and B.~Himed, ``Moving target detection using distributed
  {MIMO} radar in clutter with nonhomogeneous power,'' \emph{IEEE Trans. Signal
  Process.}, vol.~59, no.~10, pp. 4809--4820, 2011.

\bibitem{John2006Chapter4}
\emph{The Fritz John and Karush-Kuhn-Tucker Optimality Conditions}.\hskip 1em
  plus 0.5em minus 0.4em\relax John Wiley \& Sons, Ltd, 2006, ch.~8, pp.
  343--468.

\bibitem{Lehmann2006}
N.~H. Lehmann, A.~M. Haimovich, R.~S. Blum, and L.~Cimini, ``High resolution
  capabilities of {MIMO} radar,'' in \emph{2006 Fortieth Asilomar Conf.
  Signals, Syst. Comput.}, 2006, pp. 25--30.

\bibitem{Zhang2022Survey}
J.~A. Zhang, M.~L. Rahman, K.~Wu, X.~Huang, Y.~J. Guo, S.~Chen, and J.~Yuan,
  ``Enabling joint communication and radar sensing in mobile networks—a
  survey,'' \emph{IEEE Commun. Surveys {\&} Tutorials}, vol.~24, no.~1, pp.
  306--345, 2022.

\bibitem{Sakhnini2022}
A.~Sakhnini, S.~De~Bast, M.~Guenach, A.~Bourdoux, H.~Sahli, and S.~Pollin,
  ``Near-field coherent radar sensing using a massive {MIMO} communication
  testbed,'' \emph{IEEE Trans. Wireless Commun.}, vol.~21, no.~8, pp.
  6256--6270, 2022.

\bibitem{MatrixCookbook2012}
K.~B. Petersen and M.~S. Pedersen, ``The matrix cookbook,'' nov 2012.

\bibitem{Zhang2005Schur_complement}
F.~Zhang, \emph{The Schur Complement and Its Applications}.\hskip 1em plus
  0.5em minus 0.4em\relax New York: Springer New York, NY, 2005.

\bibitem{Mitrinovic1993}
D.~S. Mitrinovi{\'{c}}, J.~E. Pe{\v{c}}ari{\'{c}}, and A.~M. Fink,
  \emph{Cauchy's and Related Inequalities}.\hskip 1em plus 0.5em minus
  0.4em\relax Dordrecht: Springer Netherlands, 1993, pp. 83--98.

\bibitem{Mitrinovic1993Tri}
------, \emph{Triangle Inequalities}.\hskip 1em plus 0.5em minus 0.4em\relax
  Dordrecht: Springer Netherlands, 1993, pp. 473--513.

\bibitem{Ouyang2016}
X.~Ouyang and J.~Zhao, ``Orthogonal chirp division multiplexing,'' \emph{IEEE
  Trans. Commun.}, vol.~64, no.~9, pp. 3946--3957, 2016.

\end{thebibliography}

\IEEEpeerreviewmaketitle

\end{document}